\documentclass[twocolumn,tighten]{aastex63}
\hypersetup{linkcolor=red,citecolor=blue,filecolor=cyan,urlcolor=magenta}

\usepackage[fleqn]{amsmath}
\usepackage{natbib}
\usepackage{amssymb}
\usepackage{txfonts}
\usepackage{graphicx}
\usepackage{multirow}
\usepackage{dcolumn}
\usepackage{array}
\usepackage{xcolor}
\usepackage{float}
\usepackage{bm}

\received{February 06, 2021}
\revised{March 22, 2021}
\accepted{April 01, 2021}
\shorttitle{Thermophysical model for realistic surface layers on airless small bodies}
\shortauthors{Yu, \& Ip}

\begin{document}
\title{Thermophysical model for realistic surface layers on airless small bodies: applied to study the spin orientation and surface dust properties of (24) Themis from WISE/NEOWISE multi-epoch thermal lightcurves}

\correspondingauthor{Liang-Liang Yu}
\email{yullmoon@live.com; lilyu@must.edu.mo}

\author{Liang-Liang Yu}
\affiliation{State Key Laboratory of Lunar and Planetary Sciences, Macau University of Science and Technology, Macau, China}

\author{Wing-Huen Ip}
\affiliation{State Key Laboratory of Lunar and Planetary Sciences, Macau University of Science and Technology, Macau, China}
\affiliation{Institute of Astronomy, National Central University, Jhongli, Taoyuan City 32001, Taiwan}

\begin{abstract}
This work proposes a thermophysical model for realistic surface layers on airless small bodies (RSTPM), for the use of interpreting their multi-epoch thermal lightcurves (e.g WISE/NEOWISE). RSTPM considers real orbital cycle, rotation cycle, rough surface, temperature dependent thermal parameters, as well as contributions of sunlight reflection to observations,  hence being able to produce precise temperature distribution and thermal emission of airless small bodies regarding the variations in orbital time scales. Details of the physics, mathematics and numerical algorithms of RSTPM are presented. When used to interpret multi-epoch thermal lightcurves by WISE/NEOWISE, RSTPM can give constraints on the spin orientation and surface physical properties, like mean thermal inertia or mean size of dust grains, roughness fraction, albedo and so on via radiometric procedure. As an application example, we apply this model to the main-belt object (24) Themis, the largest object of the Themis family, which is believed to be the source region of many main-belt comets. We find multi-epoch (2010, 2014-2018) observations of Themis by WISE/NEOWISE, yielding 18 thermal lightcurves. By fitting these data with RSTPM, best-fit spin orientation of Themis is derived to be ($\lambda=137^\circ$, $\beta=59^\circ$) in ecliptic coordinates, the mean radius of dust grains on the surface is estimated to be $\tilde{b}=140^{+500}_{-114}(6\sim640)~\mu$m, indicating the surface thermal inertia to vary from $\sim3\rm~Jm^{-2}s^{-0.5}K^{-1}$ to $\sim60\rm~Jm^{-2}s^{-0.5}K^{-1}$ due to seasonal temperature variation. Further analysis found that thermal light curves of Themis show a weak rotation-phase dependent feature, indicative of heterogeneous thermal properties or imperfections of lightcurve inversion shape model.
\end{abstract}

\keywords{Infrared photometry --- Small Solar System bodies --- Asteroid surfaces --- Computational methods}

\section{Introduction}
In the solar system, there exist numerous small bodies, which are believed to be
small planetesimals that did not grow large enough to become planets. Thus some
small bodies, especially low-albedo asteroids (e.g. C-types) and fresh comets,
should contain primitive materials remaining from the formation of the Solar System,
and hence can provide important clues to the composition of the solar nebula in which
planets formed, thus improve our understanding of the origin of Solar system.

Temperature distributions of the surface and subsurface layers on small bodies are
crucial for the study about thermophysical properties of their surface materials.
To obtain the surface and subsurface temperature distributions, we would need a so-called
'surface thermophysical model', which aims to simulate such temperature distributions
on the basis of the realistic physical conditions, including the orbital motion, rotation
state, shape topography, surface roughness and surface thermophyical parameters (e.g
thermal inertia).

The first-generation thermophysical models (TPMs) of small bodies, typically like
\citet{Spencer1990}, \citet{Lagerros1996a, Lagerros1996b, Lagerros1997, Lagerros1998},
and \citet{Delbo2004}, are mainly designed for the so-called 'radiometric method', which
aims to interpret disk-integrated thermal emission observations of asteroids. Due to the
limitation of both spatial and time resolution of astronomical instruments at that age,
very limited thermal infrared observations could be obtained for few asteroids of interest,
causing that we could only estimate the global mean thermal inertias and mean roughnesses
of these asteroids. However, the appearance of roughness makes the surface emit in a
non-Lambertian way, causing more flux to be observed at low solar phase angles.
This effect is known as the "thermal infrared beaming effect" \citep{Lagerros1998}, which
leads to somewhat similar effect as a low thermal inertia surface. Thus parameters of
thermal inertia and roughness have inevitable degeneracy in the radiometry procedure.
To remove the degeneracy of thermal inertia and roughness, we would need thermal infrared
observations at multiple solar phase angles, namely observations at multiple epoches.

During the past thirty years, thanks to numerous new thermal infrared data of small
bodies from space telescopes --- IRAS, Spitzer, AKARI and WISE/NEOWISE, and high-precision
in-situ thermal infrared imaging from space missions of small bodies --- Rosetta,
Hayabusa2 and OSIRIS-REx, the requirements of observations at multiple epochs have been well met.
Particularly, the WISE/NEOWISE mission has obtained multi-epoch thermal lightcurves
of many small bodies. Now with these observations, we are able to remove the degeneracy
of thermal inertia and roughness, and simultaneously obtain constraints of their values
with the radiometric model. Hence, thermophysical modelling of small bodies has made
extraordinary progress in recent years. A review of previous TPMs of small bodies can be
found in \citet{Delbo2015}, and many updated version of TPMs have been proposed to be
applied to specific cases \citep{Rozitis2011,Davidsson2014,Hannus2015}.
\citet{Rozitis2011} made progress by modelling thermal emission of rough surface in
consideration of shadowing effect, scattering of sunlight and self-heating within the
rough region. \citet{Davidsson2014} made improvements by considering 3D heat conduction and
roughness on spatial scales smaller than the thermal skin depth. \citet{Hannus2015}
further introduced a varied shape TPM scheme, where asteroid shape and pole uncertainties
are taken into account.

With these updated TPMs, the "thermal infrared beaming effect" on airless bodies (e.g.
Moon, small satellite, asteroids and even comets) can be well explained, and thus the
degeneracy of thermal inertia and roughness can be well resolved from the radiometric
procedure. However, there still remains problems when we have to explain multi-epoch
thermal infrared data from WISE/NEOWISE with the current TPMs:

First, for some small bodies, especially main-belt asteroids, W1-band and W2-band observations
of WISE/NEOWISE could contain a significant amount of reflected sunlight. Sunlight
reflection is related to the surface albedo and roughness, which also have important
influence on the surface thermal emission. How to model sunlight reflection and thermal emission
of the surface with unified geometry and physical parameters becomes an important problem.

Second, if a target small body has a large orbital eccentricity or an obliquity close to 90 degrees,
it would show significantly different temperature at different epoches, such as (3200) Phaethon in
\citet{Yu2019} and (349) Dembowska in \citet{Yu2017}. Since thermal parameters including specific
heat capacity $c$ and thermal conductivity $\kappa$ are strong functions of temperature, the value
of thermal inertia defined as $\Gamma=\sqrt{\rho c\kappa}$ ($\rho$ means density) is certainly a
strong function of temperature as well. Consequently, such small bodies can have significantly
different thermal inertias at different epochs \citep[see, for example, ][]{Rozitis2018}. In such
cases, a mean thermal inertia derived by the radiometric procedure may not well represent the
thermophysical properties the surface materials.

If there is a dust mantle (regolith layer) on the surface, the specific heat capacity of the
dust mantle can be expressed as a function of temperature and material type, while the thermal
conductivity can be described as a function of temperature, porosity of dust mantle, density
and mean size of dust grains \citep{Gundlach2013}. The dust-mantle porosity and dust grain
density for small bodies with the same spectral would not be expected to differ very much
\citep{Britt2002}, whereas the mean grain size may be obviously different for various small
bodies even if they have similar spectral type. Besides, the mean size of dust grains would
be nearly unchanged at each observation epoch. Therefore, for small bodies covered by dust
mantle, it may be more appropriate to use mean size of dust grains rather than mean thermal
inertia as the free parameter in radiometric procedure.

In this paper, we propose a thermophysical model for realistic surface layers on airless
small bodies (RSTPM for short). The model considers not only real shape and rough surface,
but also real orbital cycle, rotational cycle, and even temperature dependent thermal parameters
in the thermal simulation process, as well as contribution of sunlight reflection in the infrared
radiometric procedure.
In comparison to previous models, RSTPM differs in three aspects, including: (1) A different
mathematical technique is used to solve the influence of surface roughness on the energy balance
equation of surface boundary; (2) For the aim to remove the degeneracy of thermal inertia and roughness
by interpreting multi-epoch thermal light-curves, variation of thermal parameters due to temperature
variation caused by orbital cycle and rotation cycle is taken into consideration; (3) A combination
model of simultaneously computing thermal emission and sunlight-reflection under the same surface
topography is proposed to fit infrared data in case of the data containing significant sunlight reflection.
The structure of the paper is as follows: Section 2 presents the details of the physics and mathematics
of RSTPM. Section 3 gives the description of the temperature dependent thermal parameters.
Section 4 describes the radiometric procedure. In Section 5, the application of RSTPM to
interpret multi-year thermal lightcurves of (24) Themis by WISE/NEOWISE is presented.
Finally, Section 6 gives an open discussion and brief conclusion of this model.

\section{Model Description}
\subsection{Thermal Diffusion}
For any small body in space, we could imagine that, following its rotation
as well as orbital movement, the temperature distribution $T(t,\vec{r})$ all
over the small body would vary with time, which is dominated by the energy
conservation law:
\begin{equation}
\frac{\partial U}{\partial t}+\nabla\cdot\vec{q}=\sum Q_{\rm s},
\label{EC}
\end{equation}
where $U=U[T(t,\vec{r})]$ is the density of internal energy, $t$ represents time,
$\vec{r}$ means position vector, $\vec{q}$ is the heat flux, and $Q_{\rm s}$
represents possible energy production source, such as energy released
by the decay of $^{26}{\rm Al}$.

Generally, the so-called specific heat capacity $c(T)$, defined to be the amount of heat
required to raise the temperature of unit mass substance by one degree, is introduced as
\begin{equation}
c(T)=\frac{1}{\rho}\frac{\partial U}{\partial T},
\end{equation}
so that the first term in Equation (\ref{EC}) can be re-written as
\begin{equation}
\frac{\partial U}{\partial t}
=\frac{\partial U}{\partial T}\frac{\partial T}{\partial t}
=\rho c_{\rm v}(T)\frac{\partial T}{\partial t}
~{\rm or}~\rho c_{\rm p}(T)\frac{\partial T}{\partial t}~,
\end{equation}
in consideration of whether the system is under constant volume ($c_{\rm v}(T)$)
or constant pressure ($c_{\rm p}(T)$).

For small bodies, the thermal process generally happens under constant pressure.
Thus the specific heat capacity at constant pressure $c_{\rm p}(T)$ should be adopted.
In the case of an airless small body, generally no mass transfer happens, thus density
$\rho$ should be constant; and no internal heat source, the item $\sum Q_{\rm s}=0$ and
can be ignored. Then the energy conservation Equation (\ref{EC}) can be rewritten as
the general thermal diffusion equation:
\begin{equation}
\rho c_{\rm p}(T)\frac{\partial T}{\partial t}=-\nabla\cdot\vec{q}.
\label{thd}
\end{equation}

We can describe the shape of a small body with a polyhedron composed of N triangle facets,
thus the small body could been divided into numerous tiny voxels in such a way that each
voxel could be marked by two number $(m,n)$, where $m$ means the radial direction towards
surface facet $m$, and $n$ means the $n$th voxel below the facet. For each voxel,
integrating the two side of Equation (\ref{thd}) leads to the following equation:
\begin{equation}
\int\rho c(T)\frac{\partial T}{\partial t}dV=-\int\nabla\cdot\vec{q}dV=\oint\vec{q}\cdot (-d\vec{S}).
\label{gheat}
\end{equation}
If the voxel $(m,n)$ is small enough, its temperature can be assumed as constant within
the voxel space, then Equation (\ref{gheat}) can be discretized as follows:
\begin{equation}
\begin{aligned}
\rho c(T)\frac{\delta T}{\delta t}V_{mn}
=&\sum_{\alpha}\vec{q}_{\alpha}\cdot\vec{S}_{\alpha} \\
=&\vec{q}_{\perp\uparrow}\cdot\vec{S}_{\perp\uparrow} +
\vec{q}_{\perp\downarrow}\cdot\vec{S}_{\perp\downarrow} +
\sum_{\alpha}\vec{q}_{\parallel\alpha}\cdot\vec{S}_{\parallel\alpha}~,
\end{aligned}
\label{vthd}
\end{equation}
where $V_{mn}$ is the volume of voxel $(m,n)$, $\alpha$ stands for a possible voxel
adjoined to voxel $(m,n)$, $\vec{S}_{\alpha}$ represents the cross-section
area-vector between the two voxels, $\perp$ stands for radial conduction, and
$\parallel$ stands for lateral conduction.

Heat flow between two voxels is the result of spatial gradient of temperature:
\begin{equation}
\vec{q}=-\kappa(T)\nabla T~,
\end{equation}
where $\kappa(T)$ is the so-called thermal conductivity.
Thus the component of the heat flux between voxel $\alpha$ and voxel $(m,n)$ and
the corresponding cross-section area-vector could be expressed as
\begin{equation}
\vec{q}_\alpha=\kappa(T_{\alpha\sim mn})\frac{T_\alpha-T_{mn}}
{\delta h_{\alpha}}\vec{l}_{\alpha\rightarrow mn}~,~
\vec{S}_\alpha=S_\alpha\vec{n}_{\alpha\rightarrow mn}~
\end{equation}
respectively, where $\vec{l}_{\alpha\rightarrow mn}$ is the unit direction vector from
voxel $\alpha$ to voxel $(m,n)$, $\vec{n}_{\alpha\rightarrow mn}$ represents the
unit normal vector, $\delta h_{\alpha}$ means the average distance between
voxel $\alpha$ and voxel $(m,n)$.

Assume the typical size of a facet in the shape model we utilized is $l_{\rm facet}$,
and the typical thermal penetration depth (generally named as 'thermal skin depth')
is $l_{\rm st}$. Then we can make the following approximations:
\[\delta h_\perp \sim l_{\rm st},~S_\perp \sim  l_{\rm facet}^2,~
\delta h_\parallel \sim  l_{\rm facet},~S_\parallel \sim l_{\rm st}^2,\]
\begin{equation}
\frac{S_\parallel}{\delta h_\parallel}\sim
\left(\frac{l_{\rm st}}{l_{\rm facet}}\right)^3\frac{S_\perp}{\delta h_\perp}.
\end{equation}
For typical small bodies, $l_{\rm facet}(\sim10\rm{m})$ is far more larger than
$l_{\rm st}(\sim10^{-2}~\rm{m})$, and
\[\frac{l_{\rm st}}{l_{\rm facet}}\sim10^{-3},~~
\frac{S_\parallel}{h_\parallel}\sim10^{-9}\frac{S_\perp}{h_\perp},\]
thus the lateral conduction could be sufficiently small to be ignored.

On the other hand, the radial conduction items could be further simplified via
the following approximations
\[S_{\perp\uparrow} \sim S_{\perp\downarrow} \sim S_\perp~,\]
\[V_{mn}\sim \delta h  S_\perp~,\]
\[\vec{q}_{\perp\uparrow}\cdot\vec{S}_{\perp\uparrow}
\sim\kappa(T_{m,n})\frac{T_{m,n-1}-T_{mn}}{\delta h}S_\perp~,\]
\[\vec{q}_{\perp\downarrow}\cdot\vec{S}_{\perp\downarrow}
\sim\kappa(T_{m,n+1})\frac{T_{m,n+1}-T_{mn}}{\delta h}S_\perp~.\]
Then Equation (\ref{vthd}) could be simplified to be 1D heat conduction equation as
\begin{equation}
\rho c(T)\frac{\delta T}{\delta t}\approx
\frac{\vec{q}_{\perp\uparrow}\cdot\vec{S}_{\perp\uparrow} +
\vec{q}_{\perp\downarrow}\cdot\vec{S}_{\perp\downarrow}}{V_{mn}}
\approx\frac{\delta}{\delta h}\left[\kappa(T)\frac{\delta T}{\delta h}\right],
\label{onethd}
\end{equation}
which would be precise enough to model temperature distribution of the surface layers
of small bodies. But if $l_{\rm st}$ is comparable to $l_{\rm facet}$, 3D thermal
diffusion model would be necessary.

\subsection{Boundary Conditions}
For a voxel at the very surface of an airless small body in space, heat flow
around this voxel contains, not only conduction between adjoining voxels, but
also absorbed incident solar flux and escaped thermal emission flux, thus the
conservation of energy gives the surface boundary condition as:
\begin{equation}
\frac{\delta U}{\delta t}=Q_{\rm absorbed}-Q_{\rm emitted}+Q_{\rm conduction}=0,
\label{scoe}
\end{equation}
if considering quasi-equilibrium state.

However, for some small bodies, especially irregular-shape asteroids,
(e.g. Eros, Toutatis and so on), the effect from topography and roughness
are significant, which are necessary to be considered in the surface
boundary condition.

Topography and roughness both indicate the fluctuation of height on the surface.
Their difference lies in the spatial scale, where topography refers to a relatively
large spatial scale ($>$ tens of meters), which only causes shadow, while roughness
refers to a macroscopic small spatial scale (meters), which can cause not only shadow
but also multi-scattering of sunlight and self-heating by self-thermal emission.

Therefore, for a small body, if its 3D shape model is composed of $\sim2000$ surface
elements, the typical size of a facet element $l_{\rm facet}$ can be treated as the
spatial scale of topography, while the spatial scale of roughness, say $l_{\rm R}$,
should be much smaller than $l_{\rm facet}$, but far larger than the thermal skin
depth $l_{\rm st}$ ($l_{\rm st}\ll l_{\rm R}\ll l_{\rm facet}$).

If the topography around facet $m$ is significant, probably causing it to be
shaded by other facets, then no sunlight will shine into this facet.

If the facet $m$ is a significant rough surface, we can further divide it into
several sub-facets, where each sub-facet $i$ is treated as smooth Lambertian
surface. Then several significant effects would arise within the sub-facets,
such as:

\textbf{(a)}. Some sub-facets may be shaded, causing the temperature of these
sub-facets to be lower than other sunlit sub-facets. If the small body is observed
at zero phase angle, all observed sub-facets would be the hotter ones, but when
the observation phase angle deviates from zero, the observed amount of colder
sub-facets would significantly increase.

\textbf{(b)}. The incident sunlight can be multi-scattered among the sub-facets
visible to each other. Thus the total light flux $L(i)$ incident onto the
sub-facet $i$ includes not only the light directly from the Sun, but also
the scattering light $L_{\rm scat}$ from surrounding visible sub-facets
\begin{equation}
L(i)=L_{\rm s}v_{i}\psi_{i}+L_{\rm scat},
\label{incli}
\end{equation}
where $v_{i}=1$ indicates that the sub-facet $i$ is visible to
the Sun (otherwise $v_{i}=0$);
\begin{equation}
\psi_{i}=\max(\vec{n}_{i}\cdot\vec{n}_{\odot},0)~,
\end{equation}
in which $\vec{n}_{i}$ is the unit normal vector of sub-facet $i$, and
$\vec{n}_{\odot}$ is the unit vector pointing to the opposite direction of sunlight;
$L_{\rm s}$ is the integrated solar flux at the distance of the asteroid,
which can be approximated by
\[L_{\rm s}=\frac{L_{\odot}}{d_{\odot}^{2}}~,\]
where $L_{\odot}$ is the solar constant, about $1361.5Wm^{-2}$, and $d_{\odot}$
is the heliocentric distance in AU. As a result, the effective amount of light
energy reflected out from the whole rough surface can be reduced, leading to
an effective Bond albedo $A_{\rm eff,B}$ smaller than the $A_{\rm B}$ of the
smooth Lambertian surface.

\textbf{(c)}. The sub-facets can absorb thermal radiation $R_{\rm th}$ from
surrounding visible facets. As a result, the colder sub-facets can be heated
by the hotter ones, this effect is known as the "self-heating effect". So the
net thermal flux that escape to the space from a sub-facet $i$ can be written as
\begin{equation}
E_{\rm th}(i)=\varepsilon\sigma T_i^{4}-R_{\rm th}(i),
\end{equation}
where $T_i$ represents the temperature of the sub-facet $i$, $\varepsilon$ is
the averaged thermal emissivity over the entire emission spectrum of the smooth
Lambertian surface, $\sigma$ is the Stefan Boltzmann constant. As a result,
the net fraction of thermal emission energy that escape out to space from the whole
rough facet would be reduced.

\begin{figure}[htpb]
\includegraphics[scale=0.42]{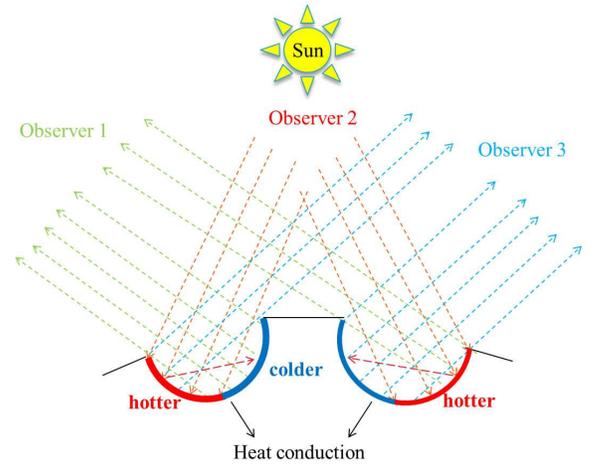}
  \centering
  \caption{Effects of shadow, multi-scattering, self-heating and beaming on rough surface.
  }\label{Rsurf}
\end{figure}

These effects together contribute to the so-called 'thermal infrared beaming effect'
\citep{Lagerros1998} --- the observed disk-integrated flux would significantly
decrease when the observation phase angle deviates from zero (Figure \ref{Rsurf}).

Therefore considering both topography and roughness, the surface boundary
condition (\ref{scoe}) for a sub-facet on rough surface can be expressed as:
\begin{equation}
(1-A_{\rm B})L(i)+R_{\rm th}(i)-\varepsilon\sigma T_i^{4}
+\kappa(T)\frac{\delta T}{\delta h}\Big|_{h=0}=0~.
\label{ssc}
\end{equation}

On the other hand, when there are no internal heat source in the small body, and the
small body is large enough so that periodical variation of temperature happens only
in the very thin surface layers during the time scale we consider, an isothermal core
would form at internal region. Thus the radial temperature gradient below a certain
depth would vanish, which gives an internal boundary condition as
\begin{equation}
\frac{\delta T}{\delta h}\Big|_{h\to\infty}\to 0 ~.
\label{inbc}
\end{equation}

\subsection{Roughness Representation}
In order to quantitatively consider the influence of roughness, we should establish
a roughness model to mathematically represent it in the surface boundary condition.
A good and widely used way to model the surface roughness \citep{Spencer1990,Lagerros1998,Rozitis2011}
is to express it by a fractional coverage of macroscopic bowl-shaped craters, symbolized
by $f_{\rm r}$ ($0\leq f_{\rm r}\leq 1$), whereas the remaining fraction, $1-f_{\rm r}$,
represents a smooth flat Lambertian surface. The configuration of the used macroscopic
crater can be described by the depth-to-diameter ratio
$\xi=h/D_{\rm rim}\geq 0$.

\begin{figure}[htbp]
\includegraphics[scale=0.68]{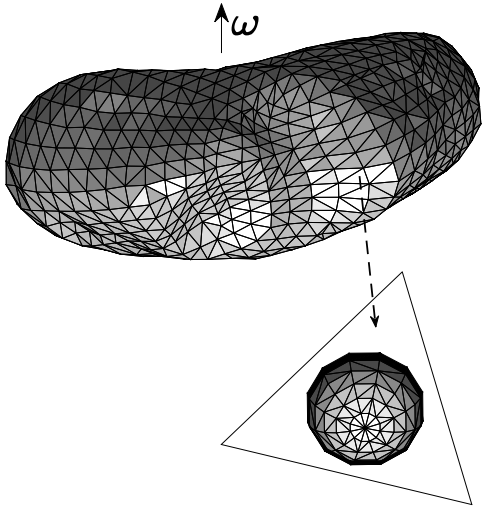}
  \centering
  \caption{The configuration for a macroscopic crater on a facet of the shape model of (433) Eros.
  The macroscopic crater is defined with $\xi=0.5$ and $\theta_{\rm rms}=50^\circ$, and is divided
  into 132 sub-facets.
  }\label{crater}
\end{figure}

The macroscopic crater can be divided into several sub-facets as Figure \ref{crater},
where each sub-facet is the same as smooth flat Lambertian surface. Then we could use
the so-called "root mean square (rms) slope" $\theta_{\rm rms}$ to measure the degree
of surface roughness. The rms slope is defined as \citep{Spencer1990}
\begin{equation}
\theta_{\rm rms}=\sqrt{\frac{\sum\limits_{i}\theta_{i}^2 a_i\cos\theta_i}{
\sum\limits_{i}a_i\cos\theta_{i}}},
\end{equation}
where $\theta_i$ is the angular slope of facet $i$ to the local horizontal surface, and
$a_{i}$ is the area of facet $i$. Generally, for a macroscopic spherical crater defined by
$\xi$, its rms slope can be uniquely calculated, like $\theta_{\rm rms}(\xi)$. Then for a
rough surface indicated by a pair of ($f_{\rm r}$,$\xi$), the RMS slope can be evaluated as
\begin{equation}
\theta_{\rm RMS}(f_{\rm r},\xi)=\sqrt{f_{\rm r}}\theta_{\rm rms}(\xi).
\end{equation}

Typically, for a hemisperical crater defined by $\xi=0.5$, the rms slop is
$\theta_{\rm rms}=50^\circ$, while considering $f_{\rm r}=0.0\sim1.0$, the
surface RMS slope $\theta_{\rm RMS}$ is in the range $0\sim50^\circ$, decided
by the roughness fraction $f_{\rm r}$.

\subsubsection{Sunlight Multi-scattering}
\begin{figure}[htbp]
\includegraphics[scale=0.35]{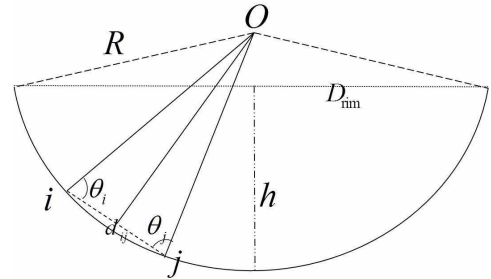}
  \centering
  \caption{The geometry of a spherical crater ($\xi=h/D_{\rm rim}$), which is
  one part of a sphere with radius $R$. Point $O$ is the centre of the sphere.
  The rim of the crater has diameter $D_{\rm rim}$, and the depth is $h$.
  }\label{crater2}
\end{figure}

Let's consider a sub-facet $i$ on the rough surface represented by
the macroscopic crater shown in Figure \ref{crater}. If the total light
flux incident onto the facet is $L(i)$, then the total light leaving the facet
can be expressed $A_{B}L(i)$, and the scattering light from surrounding visible
facets can be estimated via
\begin{equation}
L_{\rm scat}(i)=\sum_{j\neq i}f(i,j)A_{B}L(j),
\label{scat}
\end{equation}
where $f(i,j)$ is the so-called "view factor" \citep{Lagerros1998}, meaning the
fraction of radiative energy received by facet $i$ to the total radiative energy
leaving from facet $j$, assuming Lambertian energy distribution.
Therefore $f(i,j)$ can be expressed as:
\begin{equation}
f(i,j)=\left\{\begin{array}{ll}
 0, & j=i \\
 v_{i,j}a_j\frac{\cos\theta_i\cos\theta_j}{\pi d_{i,j}^2}, & j\neq i
\end{array}\right.
\end{equation}
where $v_{i,j}$ stands for the fraction of area visible to each other,
$a_j$ is the area of facet $j$, $\theta_i$ is the incidence angle on
facet $i$,  $\theta_j$ is emission angle from facet $j$, and finally
$d_{i,j}$ is  the distance between facet $i$ and $j$. Moreover, for
a bowl-shaped crater as showed in Figure \ref{crater2}, we have
\[\cos\theta_i=\cos\theta_j=\frac{0.5d_{i,j}}{R},\]
thus the view factors can be simplified to be
\begin{equation}
f(i,j)=\left\{
\begin{array}{ll}
0,& j=i \\
\frac{a_j}{4\pi R^2},& j\neq i
\end{array}\right.
\end{equation}
where $R$ is the radius of the sphere, which contains the crater.

On the other hand, when combining Equation (\ref{incli}) and
Equation (\ref{scat}), we can obtain
\begin{equation}
\begin{aligned}
L(i)=&L_{\rm s}v_{i}\psi_{i}+\sum_{j\neq i}f(i,j)A_{B}L(j) ~\Longleftrightarrow~ \\
L_{\rm s}v_{i}\psi_{i}=&L(i)-\sum_{j\neq i}f(i,j)A_{B}L(j)=\sum_{j}\textbf{A}(i,j)L(j), \\
\label{scatles}
\end{aligned}
\end{equation}
where the coefficient matrix $\textbf{A}(i,j)$ is defined as
\begin{equation}
\textbf{A}(i,j)=\left\{
\begin{array}{ll}
1, & j=i,\\
-A_{\rm B}f(i,j), & j\neq i, \\
\end{array}\right.
\label{defmA}
\end{equation}
being a symmetric matrix. Actually, Equation (\ref{scatles}) forms a system of linear equations
\begin{equation*}
\begin{array}{rcl}
\textbf{S}&=&\textbf{A}\textbf{L}~,
\end{array}
\end{equation*}
where the vector $\textbf{S}$ has elements $L_{\rm s}v_i\psi_{i}$, namely
$\textbf{S}(i)=L_{\rm s}v_i\psi_{i}$. The linear Equations can be easily
solved by Gauss Elimination Method, giving
\begin{equation*}
\begin{array}{rcl}
\textbf{L}&=&\textbf{A}^{-1}\textbf{S}~,
\end{array}
\end{equation*}
where $\textbf{A}^{-1}$ stands for the inverse matrix of $\textbf{A}$.
Therefore, the actually incident sunlight flux $L(i)$ onto the sub-facet $i$
can be obtained to be
\begin{equation}
L(i)=\sum_{j}\textbf{A}^{-1}(i,j)\textbf{S}(j)
=L_{\rm s}\sum_{j}\textbf{A}^{-1}(i,j)v_j\psi_j.
\label{Lis}
\end{equation}

\subsubsection{Thermal Self-heating}
Assume the sub-facet $i$ has temperature $T_{i}$, and assume Lambertian emission,
then the sub-facet $i$ can absorb the integrated thermal radiation
\begin{equation}
R_{\rm th}(i)=(1-A_{\rm th})\sum_{j\neq i}f(i,j)\varepsilon\sigma T_{j}^4~
\end{equation}
from surrounding visible sub-facets, where $A_{\rm th}$ is the average albedo
over the entire thermal emission spectrum, and can be related to $\varepsilon$ as
$1-A_{\rm th}=\varepsilon$ according to the Kirchhoff's law.

Therefore, the surface condition for a sub-facet on rough surface can be
further expressed as
\begin{equation}
\begin{aligned}
(1-A_{\rm B})L_{\rm s}\sum_{j}\textbf{A}^{-1}(i,j)v_j\psi_j
+&(1-A_{\rm th})\sum_{j\neq i}f(i,j)\varepsilon\sigma T_{j}^4  \\
-\varepsilon\sigma T_i^{4}
+&\kappa(T)\frac{\delta T}{\delta h}\Big|_{h=0}=0~.\\
\label{rsc}
\end{aligned}
\end{equation}

\subsubsection{Average Boundary Condition}
If integrate Equation (\ref{Lis}) for all sub-facets, the total effective
absorbed energy by the macroscopic crater can be approximated as
\begin{equation}
\begin{aligned}
(1-A_{\rm{e,B}})L_{\rm s}\psi a_{\rm rim}=&\sum_{i}(1-A_{\rm{B}})L(i)a_i \\
=&(1-A_{\rm{B}})L_{\rm s}\sum_{i}\sum_{j}\textbf{A}^{-1}(i,j)v_j\psi_ja_i,\\
\end{aligned}
\end{equation}
where $A_{\rm{e,B}}$ stands for the effective Bond albedo of the crater,
and the effective sunlit area of the crater equals
\[\psi a_{\rm rim}=\sum_{j}v_j\psi_{j}a_j.\]
Thus we can obtain
\begin{equation}
\begin{aligned}
A_{\rm{e,B}}=&1-(1-A_{\rm{B}})
\frac{\sum\limits_{i}\sum\limits_{j}\textbf{A}^{-1}(i,j)v_j\psi_ja_i}{\sum\limits_{j}v_j\psi_{j}a_j} \\
=&1-(1-A_{\rm{B}})
\frac{\sum\limits_{j}\sum\limits_{i}\textbf{A}^{-1}(j,i)v_j\psi_ja_j}{\sum\limits_{j}v_j\psi_{j}a_j} \\
\approx &1-(1-A_{\rm{B}})\left<\sum_{i}\textbf{A}^{-1}(j,i)\right>_j, \\
\end{aligned}
\end{equation}
when the area $a_i$ of the sub-facets are nearly the same. The symbol '$\langle\rangle_j$'
represents operation of averaging as
\[\left<\sum_i\textbf{A}^{-1}(i,j)\right>_j=\frac{1}{n}\sum_{j}\sum_i\textbf{A}^{-1}(i,j),\]
in which $n$ is the dimension of matrix $\textbf{A}$ and $\textbf{A}^{-1}$.

According to $\textbf{A}\textbf{A}^{-1}=\textbf{I}$ and $\textbf{A}^{\rm T}=\textbf{A}$, we have
\begin{equation}
\begin{aligned}
&\left<\sum_i\textbf{A}(i,j)\right>_j\left<\sum_{i}\textbf{A}^{-1}(j,i)\right>_j=1.  \\
\end{aligned}
\end{equation}
Besides, according to Equation (\ref{defmA})
\[\left<\sum_i\textbf{A}(i,j)\right>_j=\frac{1}{n}\sum_{j}\sum_i\textbf{A}(i,j)
=1-A_{\rm B}\frac{1}{n}\sum_{i}\sum_jf(i,j),\]
\begin{equation}
\begin{aligned}
\frac{1}{n}\sum_{i}\sum_jf(i,j)
=&\frac{1}{n}\sum_{i}\sum_{j\neq i}\frac{a_j}{4\pi R^2}
\approx\frac{1}{4\pi R^2}\sum_{j}a_j \\
=&\frac{2\pi Rh}{4\pi R^2}=\frac{4\xi^2}{1+4\xi^2}.  \\
\end{aligned}
\end{equation}
Therefore,
\[\left<\sum_{i}\textbf{A}^{-1}(j,i)\right>_j
=1\Big/\left<\sum_i\textbf{A}(i,j)\right>_j
=\frac{1}{1-A_{\rm B}\frac{4\xi^2}{1+4\xi^2}},\]
\begin{equation}
\begin{aligned}
A_{\rm{e,B}} \approx & 1-(1-A_{\rm{B}})\left<\sum_{i}\textbf{A}^{-1}(j,i)\right>_j \\
\approx & 1-\frac{(1-A_{\rm{B}})}{1-A_{\rm{B}}\left(\frac{4\xi^2}{1+4\xi^2}\right)} \\
= & \frac{A_{\rm{B}}}{1+4\xi^2(1-A_{\rm{B}})}, \\
\end{aligned}
\end{equation}
and finally the averaged effective Bond albedo of a rough surface defined by ($f_{\rm r},\xi$)
can be evaluated as
\begin{equation}
A_{\rm eff,B}=(1-f_{\rm r})A_{\rm B}+f_{\rm r}\frac{A_{\rm{B}}}{1+4\xi^2(1-A_{\rm{B}})}.
\label{ABeff}
\end{equation}

On the other hand, from the point of view of energy conservation, we could define
an effective temperature $T$ and effective thermal emissivity $\varepsilon_{\rm e}$
for the macroscopic crater via
\begin{equation}
T^4=\frac{1}{a_{\rm rim}}\sum\limits_iT_{i}^4a_i~,~
\varepsilon_{\rm e}=\frac{E_{\rm c}}{\sigma T^4a_{\rm rim}},
\end{equation}
where $E_{\rm c}$ is the net thermal emission energy escaping out from the
whole crater
\begin{equation}
E_{\rm c}=\sum_iE_{\rm  th}(i)a_i=
\sum_i\left(\varepsilon\sigma T_{i}^4-R_{\rm th}\right)a_i.
\end{equation}
Then the effective thermal emissivity $\varepsilon_{\rm e}$ can be derived to be
\begin{equation}
\begin{aligned}
\varepsilon_{\rm e} = & \left(1-\frac{\sum\limits_ia_i
\sum\limits_{j\neq i}(1-A_{\rm th})\frac{a_j}{4\pi R^2}\sigma T_{j}^4}{
\sum\limits_i\sigma T_{i}^4a_i}\right)\varepsilon \\
\approx & \left(1-\frac{(1-A_{\rm th})\sum\limits_ia_i}{4\pi R^2}\right)\varepsilon \\
= & \frac{1+4\xi^2A_{\rm th}}{1+4\xi^2}\varepsilon, \\
\end{aligned}
\end{equation}
and the averaged effective thermal emissivity of a rough surface defined
by $(f_{\rm r},\xi)$ can be evaluated as
\begin{equation}
\varepsilon_{\rm eff}=(1-f_{\rm r})\varepsilon+
f_{\rm r}\frac{1+4\xi^2A_{\rm th}}{1+4\xi^2}\varepsilon.
\label{emeff}
\end{equation}

Therefore, in a view of energy conservation for the whole rough facet, integrate
the boundary condition Equation (\ref{ssc}) for all sub-facets, and use the
averaged effective Bond albedo $A_{\rm eff,B}$ (Equation \ref{ABeff}) as well as the
averaged effective thermal emissivity $\varepsilon_{\rm eff}$ (Equation \ref{emeff}),
we could obtain the average boundary condition of the rough surface as
\begin{equation}
(1-A_{\rm eff,B})L_{\rm s}\psi-\varepsilon_{\rm eff}\sigma T^{4}+
\kappa(T)\frac{\delta T}{\delta h}\Big|_{h=0}=0~,
\label{asc}
\end{equation}
which can be applied in the cases that we do not need to distinguish the temperature
difference within the rough facet.

\subsection{Standard Transformation}
In order to simplify the solution of the above equations, it is useful
to introduce a standard transformation as follows:
\begin{equation}
u=\frac{T}{T_*},
\tau =\frac{t}{t_*},
x=\frac{h}{h_*} ,
\end{equation}
where $t_*$ represents the typical time scale of the thermal variation,
$h_*$ represents the spatial scale of thermal diffusion, $T_*$ represent
the typical temperature of the system during the thermal process.

Then, the 1D heat conduction Equation (\ref{onethd}) can be transformed as
\begin{equation}
f_c(u)\frac{\delta u}{\delta\tau}=\frac{\delta}{\delta x}
\left[f_\kappa(u)\left(\frac{\alpha_* t_*}{h_*^2}\right)\frac{\delta u}{\delta x}\right],
\label{ntd}
\end{equation}
where
\[\alpha_*=\frac{\kappa_*}{\rho c_*}~,~
f_\kappa(u)=\frac{\kappa(T)}{\kappa_*}~,~
f_c(u)=\frac{c(T)}{c_*},\]
in which $\kappa_*$ and $c_*$ are the reference constant thermal conductivity and heat
capacity introduced via
\[\kappa_*=\kappa(T_{\rm *}),c_*=c(T_{\rm *}).\]

The thermal state in the surface layers can change cyclically due to the variation of
solar insolation caused by some kind of short-term cycle of motion, for example diurnal
cycle or seasonal cycle. For such thermal cycles, generally we choose
\[t_*=\frac{1}{\omega},\]
where $\omega$ is the rotational angular frequency or orbital angular frequency,
and use the so-called thermal skin depth $l_{st*}$ as $h_*$
\begin{equation}
h_*=l_{st*}=\sqrt{\alpha_* t_*}=\sqrt{\frac{\kappa_*}{\rho c_*\omega}}.
\label{lst}
\end{equation}
And then the 1D heat conduction Equation (\ref{ntd}) can be simplified as
\begin{equation}
f_c(u)\frac{\delta u}{\delta\tau}=\frac{\delta}{\delta x}
\left[f_\kappa(u)\frac{\delta u}{\delta x}\right].
\label{tf}
\end{equation}

In realistic cases, both rotation and orbital motion can cause temperature variation.
Rotation effect is more important on low-latitude regions, especially on equatorial
regions, but none such effect on polar regions. However, if the small body has
significant orbital eccentricity and axial tilt, then seasonal thermal variation
would become important, especially for high-latitude regions, and even dominates
the temperature variation on polar regions.

\subsubsection{Diurnal cycle}
For diurnal cycle, it is convenient to use the sub-solar temperature $T_{\rm ss}$
as $T_*$
\begin{equation}
T_*=T_{\rm ss}=\left[\frac{(1-A_{\rm eff,B})L_{\rm s}}{\varepsilon\sigma}\right]^{1/4},
\end{equation}
then the boundary condition Equation (\ref{rsc}) of the rough surface can be
converted into
\begin{equation}
\begin{aligned}
&\left(\frac{1-A_{\rm B}}{1-A_{\rm eff,B}}\right)\sum_{j}\textbf{A}^{-1}(i,j)v_j\psi_j
+\sum_{j\neq i}\varepsilon f(i,j)u_{j}^4 \\
=&u_{i}^{4}-\Phi_{*}f_\kappa(u)\frac{\delta u}{\delta x}\Big|_{x=0},
\label{sbc1}
\end{aligned}
\end{equation}
where
\begin{equation}
\Phi_{*}=\frac{\Gamma_*\sqrt{\omega}}{\varepsilon\sigma T_{\rm e}^{3}}~,
\label{thpa}
\end{equation}
is the characteristic thermal parameter, in which
\[\Gamma_*=\sqrt{\rho c_*\kappa_*}\]
is the corresponding characteristic thermal inertia.

\subsubsection{Seasonal cycle}
For seasonal cycle, if the orbital period is not that long in comparison to rotation period.
Then it is easy to simulate the temperature variation considering the orbital motion and rotation
simultaneously until the surface temperature gets cyclic variation.

For the case that the orbital period is far longer than rotation period, we use the sub-solar
temperature at perihelion to define $T_*$
\[T_*=\left[\frac{(1-A_{\rm eff,B})L_{\rm sp}}{\varepsilon\sigma}\right]^{1/4},~\]
where $L_{\rm sp}$ is the incident solar flux at perihelion distance $d_{\rm sp}$.
And the surface boundary Equation (\ref{sbc1}) for each facet can be rotationally averaged
at each orbital location via:
\begin{equation}
\begin{aligned}
&\left(\frac{d_{\rm sp}}{d_\odot}\right)^2\left(\frac{1-A_{\rm B}}{1-A_{\rm eff,B}}\right)
\left<\sum_{j}\textbf{A}^{-1}(i,j)v_j\psi_j\right>
+\sum_{j\neq i}\varepsilon f(i,j)\tilde{u}_{j}^4 \\
&=\tilde{u}_{i}^{4}-\Phi_{*}f_\kappa(u)\frac{\delta \tilde{u}}{\delta x}\Big|_{x=0},
\label{sbc2}
\end{aligned}
\end{equation}
where the diurnal averaged solar insolation
\[\left<\sum\limits_{j}\textbf{A}^{-1}(i,j)v_j\psi_j\right>\]
is used, so as to compute the diurnal averaged temperature $\tilde{u}_{i}$
by simulating the seasonal variation of such diurnal-averaged temperature
for several orbital periods until reaching a stable state.

Then, for a particular orbital position, the calculated diurnal-averaged surface
temperature is used as an initial temperature to simulate the diurnal cycle of
the temperature of each facet for several rotation period. In this way, we can
obtain more precise temperature distribution, and hence more precise thermal
emission in consideration of both diurnal and seasonal effect.

\subsection{Numerical Algorithm}
The partial differential Equation (\ref{tf}) and its boundary conditions
Equation (\ref{sbc1}) or Equation (\ref{sbc2}) are very complicated coupling
equations, for which it is difficult to obtain analytic solutions, thus here
we attempt to solve it with numerical algorithm.

When we want to numerically solve a time-dependent partial differential equation,
like Equation (\ref{tf}), which would reach a cyclic stable solution, it is
suitable to use a time-marching algorithm until the final periodic-stabilized
solution. Let us use $\delta\tau$ as the time step, $\delta x$ as the spatial step
and $u_{i,j}$ to represent the current temperature of depth $j$ below facet $i$,
while $u^n_{i,j}$ represents its temperature at next time step. Then with the
Crank-Nicholson scheme, we can obtain the differential forms of Equation (\ref{tf})
as:
\begin{equation}
\begin{array}{ll}
u^n_{i,j}-u_{i,j}~
=& f_{i,j-1}(u^n_{i,j-1}-u^n_{i,j})+f_{i,j}(u^n_{i,j+1}-u^n_{i,j}) \\
&+f_{i,j-1}(u_{i,j-1}-u_{i,j})+f_{i,j}(u_{i,j+1}-u_{i,j})  \\
\label{cnthd}
\end{array}
\end{equation}
where
\begin{equation}
f_{i,j}=\frac{1}{2}\frac{\tilde{f_{\kappa}(u_{i,j})}}{{f_c(u_{i,j})}}f_{\rm o}=
\frac{1}{2}\frac{\tilde{\kappa}(u_{i,j})}{c(u_{i,j})}\frac{c_*}{\kappa_*}\frac{\delta\tau}{\delta x^2},
\end{equation}
in which
\begin{equation}
\tilde{\kappa}(u_{i,j})=\frac{\kappa(u_{i,j})V_{i,j}+\kappa(u_{i,j+1})V_{i,j+1}}{V_{i,j}+V_{i,j+1}},
\end{equation}
and $V_{i,j}$ stands for the volume of voxel $(i,j)$.
By putting items of next step $u^n$ to the left side, and items of current step $u$
to the right side, Equation (\ref{cnthd}) can be further rewritten as
\begin{equation}
\begin{aligned}
 & -f_{i,j-1}u^n_{i,j-1}+(1+f_{i,j-1}+f_{i,j})u^n_{i,j}-f_{i,j}u^n_{i,j+1} \\
=& f_{i,j-1}u_{i,j-1}+(1-f_{i,j-1}-f_{i,j})u_{i,j}+f_{i,j}u_{i,j+1}. \\
\label{tthd}
\end{aligned}
\end{equation}

For the surface boundary condition Equation (\ref{sbc1}) in diurnal cycle,
to ensure the stability as well as efficiency of calculation, we could use
a semi-implicit scheme, making it converted into
\begin{equation}
(u^{n}_{i,1})^{4}+ p_1u^{n}_{i,1}=p_2,
\label{sboundary}
\end{equation}
in which
\[p_1=\Phi_{*}\tilde{f}_{\kappa}(u_{i,1})\frac{1}{\delta x},\]
\[p_2=\Phi_{*}\tilde{f}_{\kappa}(u_{i,1})\frac{u_{i,2}}{\delta x}
+{\textstyle\left(\frac{1-A_{\rm B}}{1-A_{\rm eff,B}}\right)}
\sum_{j}\textbf{A}^{-1}(i,j)v_j\psi_j
+\sum_{j\neq i}\varepsilon f(i,j)u_{j,1}^4~.\]

The surface boundary condition Equation (\ref{sbc2}) in seasonal cycle can also
be transformed in a similar way as above.

In the process of numerical calculation, we can firstly calculate the parameter
$p_1$ and $p_2$ with the temperatures and solar insolation at current time step,
and then obtain the surface temperature of next time step
\begin{equation}
u^{n}_{i,1}=BC(p_1,p_2),
\label{surft}
\end{equation}
where $BC$ is named for the function to solve the non-linear Equation (\ref{sboundary}).
On the other hand, the internal boundary condition can be simply expressed as
\begin{equation}
u^{n}_{i,N}-u^{n}_{i,N-1}=0~.
\label{iboundary}
\end{equation}

Then Equation (\ref{tthd}), Equation (\ref{surft}), and Equation (\ref{iboundary})
could be combined into a special tridiagonal system of equations as
\begin{small}
\begin{equation}
\left[\begin{array}{cccccc}
b_1 & c_1 &     &         &          & 0 \\
a_2 & b_2 & c_2 &         &          &   \\
    & a_3 & b_3 & c_3     &          &   \\
&\multicolumn{3}{c}{.~.~.} &          &\\
    &     & a_j & b_j     & c_j      &   \\
    &    &\multicolumn{3}{c}{.~.~.}   &   \\
    &     &     & a_{N-1} & b_{N-1}  & c_{N-1} \\
0   &     &     &         & a_N      & b_N
\end{array}\right]
\left[\begin{array}{c}
u^{n}_{i,1} \\
u^{n}_{i,2} \\
u^{n}_{i,3} \\
... \\
u^{n}_{i,j} \\
... \\
u^{n}_{i,N-1} \\
u^{n}_{i,N}
\end{array}\right]
=
\left[\begin{array}{c}
d_1 \\
d_2 \\
d_3 \\
... \\
d_j \\
... \\
d_{N-1} \\
d_N
\end{array}\right]
\label{TDMA}
\end{equation}
\end{small}
where
\begin{small}
\[a_j=\left\{\begin{array}{ll}
 0, & j=1 \\
 -f_{i,j-1}, & j=2,N-1 \\
 1, & j=N
\end{array}\right.\]
\[b_j=\left\{\begin{array}{ll}
 1, & j=1 \\
 1+f_{i,j-1}+f_{i,j}, & j=2,N-1 \\
 -1, & j=N
\end{array}\right.\]
\[c_j=\left\{\begin{array}{ll}
 0, & j=1 \\
 -f_{i,j}, & j=2,N-1 \\
 0, & j=N
\end{array}\right.\]
\[d_j=\left\{\begin{array}{ll}
 BC(p_1,p_2), & j=1 \\
f_{i,j-1}u_{i,j-1}+(1-f_{i,j-1}-f_{i,j})u_{i,j}+f_{i,j}u_{i,j+1}, & j=2,N-1 \\
 0. & j=N
\end{array}\right.\]
\end{small}

Equation (\ref{TDMA}) is still a nonlinear equation, because the
coefficients contain the temperature dependent function $f_{i,j}$.
To solve this kind of nonlinear heat conduction equation, we could
use the so-called '\emph{predictor-corrector}' method, which is a
two-step iterative procedure. The advance from temperature $u$ at
current time $\tau$ to temperature $u^n$ at next time $(\tau+\delta\tau)$
is implemented through the temperature $u^{n-1/2}$ at a intermediate
time step $(\tau+\delta\tau/2)$.

Firstly, the \emph{predictor} step, we can obtain the intermediate
temperature $u^{n-1/2}$ from the current temperature $u$ by taking
the coefficient $f_{i,j}(u)$ with current temperature $u$.

Secondly, the \emph{corrector} step, the coefficients $f_{i,j}(u^{n-1/2})$
are derived by the temperature $u^{n-1/2}$ at time step $(\tau+\delta\tau/2)$,
so as to obtain $u^n$ from $u$.

Both the \emph{predictor} and \emph{corrector} step need the solution of the
linear tridiagonal system of equations, which already has known solution algorithms.

\section{Thermal Parameters}
\subsection{Specific heat capacity}
The specific heat capacity of simple materials as a function of temperature can be
described by the Debye's theory when the temperature is not very high (lower than
the so-called Debye characteristic temperature). But for materials on planetary
surface or minerals, the relation between specific heat capacity and temperature
actually deviates from the Debye's theory. Nevertheless, the positive correlation
that specific heat capacity increases with the increasing of temperature is similar
to the Debye's formula. Hence, the Debye's formula can serve as an approximation for
the temperature dependence of specific heat capacity of minerals or planetary surface
materials.

Considering that the Debye's formula of specific heat capacity is a time-consuming
integration formula, here we present a simplified formula to approximate the positive
correlation between specific heat capacity and temperature based on the Debye's formula:
\begin{equation}
c_{\rm v}(T)=\frac{3k_{\rm B}}{m_{\rm a}}\left\{\frac{a\Big(\frac{T}{T_{\rm D}}\Big)^3
\Big[a\Big(\frac{T}{T_{\rm D}}\Big)^3+2b\Big(\frac{T}{T_{\rm D}}\Big)^2+3c\frac{T}{T_{\rm D}}+4\Big]}{
\Big[a\Big(\frac{T}{T_{\rm D}}\Big)^3+b\Big(\frac{T}{T_{\rm D}}\Big)^2+c\frac{T}{T_{\rm D}}+1\Big]^2}
\right\}~,
\label{cvt}
\end{equation}
where the coefficients
\[a\approx39.09~,b\approx14.46~,c\approx3.304,\]
$k_{\rm B}$ is the Boltzmann's constant, $m_{\rm a}=\rho/n$ is the average atom mass
of the material, and $T_{\rm D}$ is a "characteristic temperature" of the material
that would be measured by experiments.

While we have no exact information about surface materials on small bodies, we may
find their most spectral-resembled chondrites to estimate their heat capacities.
The average atom mass $m_{\rm a}$ of several chondrites are listed in Table \ref{chmas}.
\begin{table}[htbp]
 \centering
 \renewcommand\arraystretch{0.8}
 \caption{The average atom mass $m_{\rm a}$ of known chondrites according to
 composition of chondrites given by \citet{Wasson1988}.}
 \label{chmas}
 \begin{tabular}{@{}cc@{}}
 \hline
 chondrite  &  $m_{\rm a}$ \\
 Type & ($1.66053873\times10^{-27}$ kg) \\
 \hline
 CI  & 21.5515 \\
 CM  & 22.7817 \\
 H   & 24.9229 \\
 L   & 23.6574 \\
 LL  & 23.4194 \\
 EH  & 25.9846 \\
 EL  & 25.1344 \\
 \hline
\end{tabular}
\end{table}

In principle, the characteristic temperature $T_{\rm D}$ of each chondrite can
be obtained by fitting experimental data of specific heat capacities with
Equation (\ref{cvt}). By comparing Equation (\ref{cvt}) with the experimental
data of \citet{Macke2016,Opeil2016}, we suggest a constant
\[T_{\rm D}\approx700~{\rm K}\]
for various chondrites.

When the temperature gets very high, the Debye's theory is no longer suitable.
\citet{Waples2004} presents empirical formula for the specific heat capacities
of minerals and nonporous rocks at high temperature, where a relative heat
capacity is given as
\begin{equation}
\begin{aligned}
N_{c\rm p}(T)=&0.716+1.72\times10^{-3}(T-273.4) \\
&-2.13\times10^{-6}(T-273.4)^2 \\
&+8.95\times10^{-10}(T-273.4)^3, \\
\end{aligned}
\end{equation}
and the specific heat capacity is then estimated via
\begin{equation}
c_{\rm p}(T)=c_{\rm p}(T_0)\frac{N_{c\rm p}(T)}{N_{c\rm p}(T_0)}.
\end{equation}

\begin{figure}[htbp]
\includegraphics[scale=0.58]{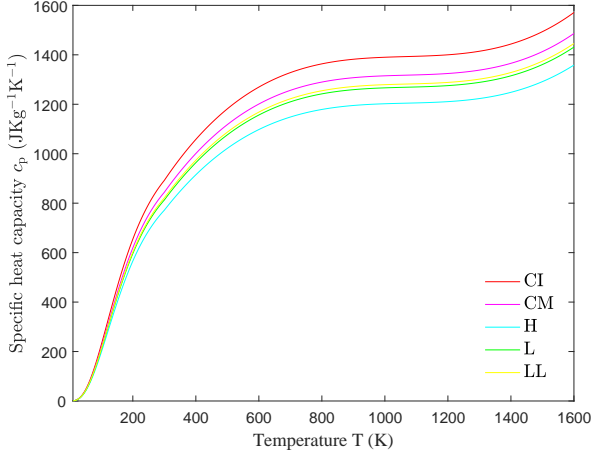}
  \centering
  \caption{Model results of specific heat capacities of CI, CM, H, L, LL chondrites.
  }\label{chcpt}
\end{figure}

Then the variation of specific heat capacity in a very large temperature scale can
be expressed as
\begin{equation}
c_{\rm p}(T)=\left\{
\begin{array}{ll}
c_{\rm v}(T), & T<300 K, \\
\\
c_{\rm v}(300)\frac{N_{c\rm p}(T)}{N_{c\rm p}(300)}, & T\geq300 K.
\end{array}\right.
\label{cpt}
\end{equation}
Here in Figure \ref{chcpt}, we present the specific heat capacities of
CI, CM, H, L, LL chondrites obtained by the above methods.

\subsection{Thermal conductivity}
For small bodies covered by dust mantle, the thermal conductivity $\kappa$ of the dust mantle
can be related to the temperature $T$, mean radius $b$ of grains and dust mantle porosity $\phi$
via the model of \citet{Gundlach2013}:
\begin{equation}
\begin{aligned}
\kappa(T,b,\phi)=&\kappa_{\rm solid}
\left(\frac{9\pi}{4}\frac{1-\mu^{2}}{E}\frac{\gamma(T)}{b}\right)^{1/3}\chi f_{1}e^{f_{2}(1-\phi)} \\
&+8\sigma\epsilon T^{3}\frac{e_{1}\phi}{1-\phi}b ~.
\label{ktrphi}
\end{aligned}
\end{equation}
The details of Equation (\ref{ktrphi}) are described in \citet{Gundlach2013}.

\begin{figure}[htbp]
\includegraphics[scale=0.58]{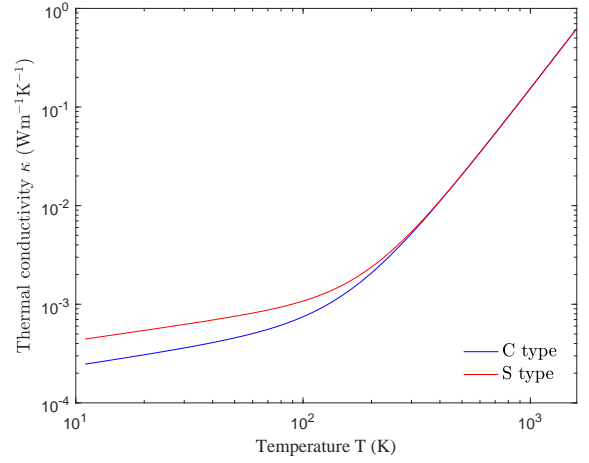}
  \centering
  \caption{Temperature dependent thermal conductivities of C-type and S-type asteroids,
  assuming mean grain radius $b=0.5~{\rm mm}$, porosity $\phi=0.5$.
  }\label{kpt}
\end{figure}

Based on Equation (\ref{ktrphi}), in Figure \ref{kpt}, we show how the thermal
conductivities change with temperature for C-type and S-type asteroids,
assuming mean grain radius $b=0.5~{\rm mm}$, porosity $\phi=0.5$.

\subsection{Thermal inertia}
\begin{figure}[htbp]
\includegraphics[scale=0.58]{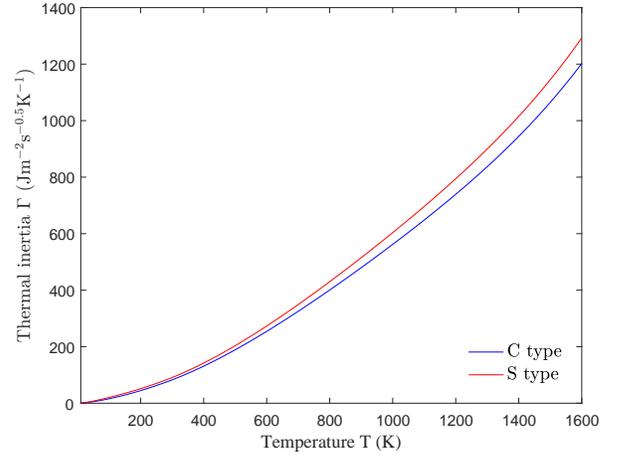}
  \centering
  \caption{Temperature dependent thermal inertias of C-type(CM) and S-type (LL) asteroids,
  assuming mean grain radius $b=0.5~{\rm mm}$, surface porosity $\phi=0.5$, grain density
  $3100~{\rm kgm^{-3}}$ for C-type but $3700~{\rm kgm^{-3}}$ for S-type.
  }\label{tht}
\end{figure}

As shown in Figure \ref{chcpt} and Figure \ref{kpt}, both the specific heat capacities
and thermal conductivities are strong functions of temperature, thus the so-called
thermal inertia
\begin{equation}
\Gamma=\sqrt{\rho c_{\rm p}\kappa},
\label{defgma}
\end{equation}
is also strongly temperature dependent (seeing Figure \ref{tht}).

According to the surface boundary condition Equation ({\ref{sbc1}}), the thermal
parameter that relates to thermal inertia decides how the surface temperature
changes. The thermal inertia, as indicated by its name, generally implies the
ability to maintain the thermal state. Thus, the larger thermal inertia is,
the slower temperature variation would be, and probably causing significant
thermal-delay effect, generating asymmetric temperature distribution between
the sunrise side and the sunset side.

On the other hand, due to the strong positive correlation between thermal inertia
and temperature, the rise of temperature could increase thermal inertia, and thus
slow down the temperature increase and enhance the thermal delay effect, while
inversely, the decline of temperature would reduce thermal inertia, and thus promote
the temperature decrease. As a result, temperature distribution between the sunrise
and the sunset sides would tend to be even more asymmetric due to the temperature
dependence of the thermal parameter.

In the following sub-sections, we will use RSTPM to show how the temperature
dependence of thermal inertia enhances the asymmetry of temperature distribution
in the diurnal cycle of small bodies. As examples, let's consider S-type asteroids
at $1~{\rm AU}$ from the Sun and physical parameters listed in Table \ref{testb}.
\begin{table}[htbp]
 \centering
 \renewcommand\arraystretch{0.8}
 \caption{Assumed parameters for the test asteroids.}
 \label{testb}
 \begin{tabular}{@{}ll@{}}
 \hline
 Properties  &  Value \\
 \hline
 Heliocentric distance            & 1 AU \\
 Rotation obliquity               & $0^\circ$ \\
 Rotation period $P_{\rm R}$      & 10 hr \\
 Roughness fraction $f_{\rm r}$   & 0.0   \\
 Bond albedo $A_{\rm B}$          & 0.04  \\
 Thermal emissivity $\varepsilon$ & 0.9   \\
 Resembled chondrite              & LL    \\
 Material density $\rho_{\rm m}$  & $3700~{\rm kgm^{-3}}$ \\
 Surface porosity $\phi$          & 0.5   \\
 \hline
\end{tabular}
\end{table}

\subsubsection{Mean Thermal inertia vs Mean grain radius}
As a comparison to the realistic case, firstly we have to define a constant mean
thermal inertia, which ignores the temperature dependence. Therefore, we need a
mean temperature $\tilde{T}$, and the mean thermal inertia is defined as
\begin{equation}
\tilde{\Gamma}=\sqrt{\rho c_{\rm p}(\tilde{T})\kappa(\tilde{T})}.
\end{equation}

For the test asteroids with above conditions, the diurnal mean temperature on the
equator can be estimated to be
\[\tilde{T}=\left(\frac{(1-A_{\rm B})L_\odot}{\varepsilon\sigma\pi}\right)^{1/4}
\approx300~{\rm K}.\]

\begin{figure}[htbp]
\includegraphics[scale=0.58]{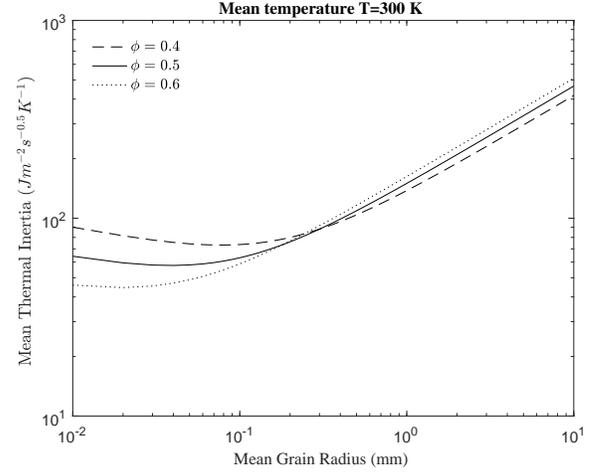}
  \centering
  \caption{The relation of mean thermal inertia and mean grain radius under the
  condition given in Table \ref{testb}
  }\label{thgr}
\end{figure}

Then the mean grain radius $\tilde{b}$ would be the main parameter that decides the
mean thermal inertia $\tilde{\Gamma}$ as shown in Figure \ref{thgr}, which shows that
larger grain radius generally means larger thermal inertia.

\subsubsection{Enhanced Asymmetric Temperature Distribution}
\begin{figure}[htbp]
\includegraphics[scale=0.58]{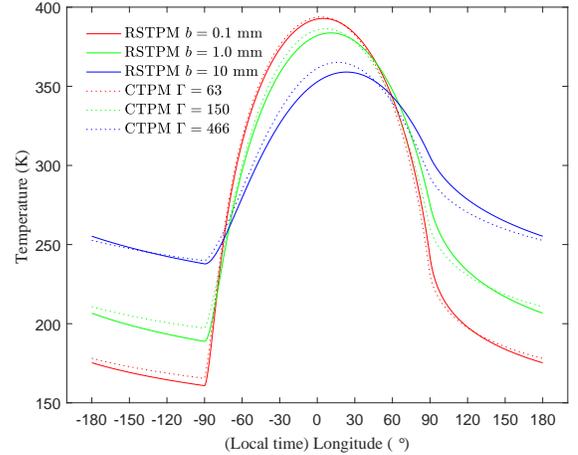}
  \centering
  \caption{Modeled diurnal temperature variation with two different models: RSTPM (solid curves),
  the model presented in this work, considers temperature dependent parameters when mean grain
  radius $b=0.1,1,10\rm~mm$ respectively; CTPM (dotted curves), a common used thermophysical model,
  ignoring temperature dependence, considers constant thermal inertia
  $\Gamma=63,150,466\rm~Jm^{-2}s^{-0.5}K^{-1}$ (red solid curve) respectively.
  }\label{tstdv}
\end{figure}

In Figure \ref{tstdv}, we show the diurnal temperature variation obtained
by two models with different thermal parameters. The solid curves (with color
in red, green and blue) are obtained by RSTPM, considering temperature dependent
thermal parameters in three cases of mean grain radius
\[\tilde{b}=0.1,1,10\rm~mm\]
respectively, while the dotted curve with the same color are obtained by Commonly
used Thermophysical Models (using CTPM for short) that ignores temperature dependence,
using the corresponding mean thermal inertia
\[\tilde{\Gamma}=63,150,466\rm~Jm^{-2}s^{-0.5}K^{-1}\]
respectively.

Firstly, if we do comparisons between curves with different color, we can see
that larger thermal inertia tends to generate stronger thermal delay effect or
asymmetric temperature distribution.

Secondly, if we do comparisons between the solid curves and the corresponding
dotted curves with the same color, the temperature distribution tends to be
more asymmetric if thermal parameters' temperature dependence is taken into
consideration. Such effect can lead to a temperature difference as large as
$\sim10$ K, which can hence induce variation of thermal emission within
\[\sim\left(\frac{300+10}{300}\right)^4\sim14\%.\]

Finally, the asymmetry enhancement tends to be more significant for cases with
larger thermal inertia (blue curves).

\section{Thermal infrared Radiometry}
The infrared radiation from a small body is related to its size as well as temperature distribution,
which is decided by the surface thermophyscial properties (e.g. albedo, roughness and thermal inertia).
Thus these properties could be well determined by fitting the measurements of its infrared radiation
with the surface thermophysical model. This procedure is known as the so-called "Thermal Infrared
Radiometry". However, for any body except the Sun in the solar system, the observed infrared radiation
is the integration of its own thermal emission and the reflected sunlight, especially radiation at
wavelength $<5~\mu$m, containing significant fraction of sunlight (e.g. W1 and W2 band of WISE/NEOWISE,
see Figure \ref{rlfraction} in Section 5.4). Thus both thermal emission and sunlight reflection should
be taken into account in the radiometric model.

\subsection{Disk-integrated thermal emission}
On the basis of the previously described roughness model, we treat the facet $i$ of the
shape model and the sub-facet $j$ in the crater on facet $i$ both to be smooth Lambertian
surface, and hence thermal radiation from them can be approximated by Lambertian radiation.
So for a given epoch with a certain observation phase angle $\alpha$ and distance $\Delta$,
thermal emission from facet $i$ and sub-facet $ij$ that can be observed by the telescope
will be $\epsilon(\lambda)\pi B_if_i$ and $\epsilon(\lambda)\pi B_{ij}f_{ij}$, where
$\epsilon(\lambda)$ is the monochromatic emissivity at wavelength $\lambda$, $f_i$ and
$f_{\rm ij}$ are the view factors of facet $i$ and sub-facet $ij$ relative to the telescope,
$B_i$ and $B_{ij}$ are Planck intensity function on a temperature $T_{i}$ and $T_{ij}$
\begin{equation}
B(\lambda, T_{i})=\frac{2hc^{2}}{\lambda^{5}}\frac{1}{\exp\big(\frac{hc}{\lambda kT_{i}}\big)-1}.
\end{equation}
The so-called view factor $f_i$ is defined as
\begin{equation}
f_i=v_i a_i\frac{\vec{n}_{i}\cdot\vec{n}_{\rm obs}}{\pi\Delta^{2}},
\end{equation}
where $a_i$ and $\vec{n}_{i}$ are the area and normal vector of facet $i$, $\vec{n}_{\rm obs}$
is the unit vector of the telescope's direction in the body-fix coordinate system, $v_i=1$
indicates that facet $i$ is observable from the telescope, otherwise $v_i=0$.

With the temperature distribution $T_{i}$ and $T_{ij}$ computed from the above numerical method,
the observable thermal emission $F_{\rm th}$ of the entire small body can be expressed as the
integration of thermal emission from both the smooth and rough surface:
\begin{equation}
F_{\rm th}(\lambda)=(1-f_{\rm r})\sum^{N}_{i=1}\epsilon(\lambda)\pi B_if_i+
f_{\rm r}\sum_{i}^{N} \sum_{j}^{M}\epsilon(\lambda)\pi B_{ij}f_{ij}.
\label{thflux}
\end{equation}

\subsection{Disk-integrated sunlight reflection}
While it could be a good approximation to calculate the absorption of solar energy
by assuming Lambertian surface, such approximation will not be good enough if we care
about the exact reflected sunlight. Actually, sunlight reflection by realistic planetary
surface deviates largely from the ideal Lambertian reflection, due to multiple effects
including asymmetric scattering by irregular-shape dust particles and macroscopic-roughness
(far larger than dust particle size) induced beaming effect. The macroscopic roughness model
can be similar to the thermal roughness model above. To consider asymmetric scattering,
we use the combined Lambert-Lommel-Seeliger law that introduce a correction coefficient
$C_{\rm L}$ to the Lambertian reflection as
\begin{equation}
C_{\rm L}(\psi_i,\psi_{\rm o,i},\alpha,w_{\rm f})=f(\alpha)\left(w_{\rm f}+\frac{1}{\psi_i+\psi_{o,\rm i}}\right),
\label{ScatCoeff}
\end{equation}
where $\psi_i$ and $\psi_{\rm o,i}$ are the cosines of the incident angle and emergence angle
on facet $i$ respectively, $\alpha$ is the solar phase angle; $f(\alpha)$ is the phase correction
function, according to \citep{Kaasalainen2001b},
\[f(\alpha)\sim0.5\exp(-\alpha/0.1)-0.5\alpha+1.\]
Parameter $w_{\rm f}$ represents the weight of Lambertian term in the scattering law, so we name
it "scattering weight-factor". To ensure $0\leq C_{\rm L}\leq1$, the scattering weight-factor
$w_{\rm f}$ is required to be $0\leq w_{\rm f}\leq0.5$. The value $w_{\rm f}$ can be determined
by doing optimization fitting to observations of sunlight reflection.

Hence, for a given epoch with a certain observation phase angle $\alpha$ and distance $\Delta$,
the reflection of sunlight from facet $i$ and sub-facet $ij$ can be expressed as:
\begin{equation}
F_{\rm rl,i}(\lambda)=\pi B(\lambda,5778)\frac{R_{\rm sun}^2}{r_{\rm helio}^2}\cdot A_{\rm b}(\lambda)\cdot\psi_i\cdot f_i\cdot C_{\rm L},
\end{equation}
\begin{equation}
F_{\rm rl,ij}(\lambda)=\pi B(\lambda,5778)\frac{R_{\rm sun}^2}{r_{\rm helio}^2}\cdot A_{\rm b}(\lambda)\cdot\psi_{ij} \cdot f_{ij}\cdot C_{\rm L},
\end{equation}
where $\psi_i$ and $\psi_{ij}$ are the cosine values of the solar altitudes, $R_{\rm sun}$ is
the radius of the Sun, $r_{\rm helio}$ nearly equals to the heliocentric distance of the asteroid,
$B(\lambda, T)$ is the Planck intensity function, and $A_{\rm b}(\lambda)$ is the albedo at
wavelength $\lambda$.

Then the total reflected sunlight that can be observed by the telescope is the integration
of reflection from all observable facets:
\begin{equation}
\begin{split}
\label{refl}
F_{\rm rl}(\lambda)=(1-f_{\rm r})\sum_{i}^{N}F_{\rm rl,i}+f_{\rm r}\sum_{i}^{N} \sum_{j}^{M}F_{\rm rl,ij}.
\end{split}
\end{equation}
And the total radiation flux that can be observed by the telescope at wavelength $\lambda$
would be the sum of thermal emission and sunlight reflection:
\begin{equation}
F_{\rm model}(\lambda)=F_{\rm th}(\lambda)+F_{\rm rl}(\lambda).
\label{fmodel}
\end{equation}

\subsection{Thermal Infrared Beaming effect}
As mentioned in section 2.2, the appearance of roughness could lead to the so-called
thermal infrared beaming effect, making the observed disk-integrated emission flux
to decrease significantly when the observation phase angle deviates from zero. Here
we use RSTPM to investigate the beaming effect. As an example, a testing asteroid
with parameters listed in Table \ref{testb} and two cases of roughness fraction
$f_{\rm r}=0$ and 1.0 are investigated.

\begin{figure}[htbp]
\includegraphics[scale=0.58]{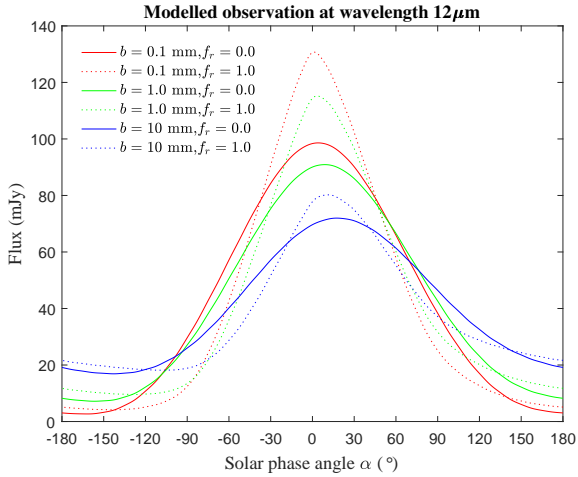}
  \centering
  \caption{Modelled $12~\mu m$ observation flux in a equatorial view but at various
  solar phase angle for a test asteroid with parameters listed in Table \ref{testb}.
  The observation distance is fixed at $1~\rm AU$, but two roughness fraction
  $f_{\rm r}=0$ and 1.0 and three grain radius $b=$0.1, 1, and 10 mm are taken into account.
  }\label{fxsph}
\end{figure}

In Figure \ref{fxsph}, we show the modelled $12~\mu m$ observation flux in a equatorial
view but at various solar phase angle. The observation distance is fixed at $1~\rm AU$,
but different grain radius are taken into account.

Firstly, the difference between the dotted curve and the solid curve with the same
color clearly reveals the mentioned beaming effect --- more emission at zero solar
phase angle.

Secondly, the beaming effect tends to be more significant for the cases of low-thermal
inertia cases (smaller grain size). The observed flux at zero solar phase can increase
up to $\sim30\%$ (compare the red dotted curve and the red solid curve) due to the beaming
effect generated by a totally rough surface (roughness fraction = 1.0).

Thirdly, both the decreasing of thermal inertia and the increasing of roughness can lead
to the tendency of emitting more photons at zero solar phase angle. These two effects can
be degenerate for observations at insufficient solar phase angles. Therefore, in order
to reduce the degeneracy of the surface thermal inertia and roughness from thermal
infrared observations, observations at various solar phase angles are necessary.

\subsection{Radiometric Procedure}
In order to reproduce the disk-integrated infrared observation of a small body with
the RSTPM, we need its 3D shape model, rotation parameters (rotation period and
rotation axis orientation), effective diameter $D_{\rm eff}$, bond albedo $A_{\rm B}$,
emissivity, and thermophysical parameters $\rho$, $c_{\rm p}$, $\kappa$.

\subsubsection{Shape Model and Spin orientation}
The 3D shape model and spin state can be constructed by the light-curve inversion method
developed by \citet{Kaasalainen2001} if we have observed enough light-curves, or by inversion
of radar Delay-Doppler images \citep{Ostro2002}. Since observations of radar Delay-Doppler
images are more difficult than light-curve observations, light-curve inversion method has
become the most common-used method to derive shape and spin state of most small bodies.
However, even for light-curve observation, it is difficult to obtain enough light-curves
to constrain a unique spin orientation for many small bodies. It is fortunate that
WISE/NEOWISE have observed lots of thermal lightcurves of many small bodies, among which
spin orientation still has different solutions. Therefore, by interpreting the thermal
lightcurves of WISE/NEOWISE, it is possible to constrain the spin orientation to a unique
solution.

\subsubsection{Size and Albedo}
According to \citet{Fowler1992}, an asteroid's effective diameter $D_{\rm eff}$,
defined by the diameter of a sphere with the same area to that of the shape
model, can be related to its geometric albedo $p_{v}$ and absolute visual
magnitude $H_{v}$ via:
\begin{equation}
D_{\rm eff}=\frac{1329\times 10^{-H_{v}/5}}{\sqrt{p_{v}}}~(\rm km) ~.
\label{Deff}
\end{equation}
In addition, the geometric albedo $p_{v}$ is related to the effective
Bond albedo $A_{\rm eff,B}$ by
\begin{equation}
A_{\rm eff,B}=p_{v}q_{\rm ph}~,
\label{aeffpv}
\end{equation}
where $q_{\rm ph}$ is the phase integral that can be approximated by
\begin{equation}
q_{\rm ph}=0.290+0.684G~,
\label{qph}
\end{equation}
in which $G$ is the slope parameter in the $H, G$ magnitude system of
\citet{Bowell}, which can be obtained by photometric observation.

\subsubsection{Roughness Fraction}
On the other hand, the asteroid's effective Bond albedo is the averaged
result of both the albedo of smooth and rough surface, which can be expressed
as the following relationship according to Equation (\ref{ABeff})
($\xi=0.5$ for hemispherical crater):
\begin{equation}
A_{\rm eff,B}=(1-f_{\rm r})A_{B}+f_{\rm r}\frac{A_{B}}{2-A_{B}}~,
\label{abfr}
\end{equation}
where $A_{B}$ is the Bond albedo of smooth Lambertian surface. Thus an input
roughness fraction $f_{\rm r}$ and geometric albedo $p_{\rm v}$ can lead to
an unique Bond albedo $A_{B}$ and effective diameter $D_{\rm eff}$ to be used
to fit the observations.

\subsubsection{Thermal emissivity}
We can use the bond albedo $A_{B}$ to approximate the reflectance $A_{\rm b}(\lambda)$
at an observation wavelength $\lambda$, so as to calculate the reflect sunlight at
wavelength $\lambda$. On the other hand, according to Kirchhoff's law, the monochromatic
emissivity $\epsilon(\lambda)$ at wavelength $\lambda$ can be approximatively related to
$A_{\rm b}(\lambda)$ via
\[\epsilon(\lambda)=1-A_{\rm b}(\lambda),\]
thus enabling the computation for the thermal emission at wavelength $\lambda$.
Under such approximation, the size, albedo and emissivity are related to each other,
thus becoming one free parameter in the fitting procedure.

\subsubsection{Mean Thermal inertia}
For the thermophysical parameters, as a first approximation, we may ignore their
temperature dependence, and assume a mean thermal inertia of the whole surface to
obtain a corresponding mean thermal parameter
\begin{equation}
\tilde{\Phi}=\frac{\tilde{\Gamma}\sqrt{\omega}}{\varepsilon\sigma T_{\rm e}^{3}}~,
\end{equation}
then we are able to obtain the surface temperature distribution to fit the thermal
infrared observations.

Thus we actually have three free parameters --- mean thermal inertia, roughness fraction,
and geometric albedo (or effective diameter) that would be extensively investigated
in the fitting process. We use the so-called reduced $\chi^{2}_{\rm r}$ defined as
\begin{equation}
\chi^{2}_{\rm r}=\frac{1}{n-3}\sum^{n}_{i=1}
\Big[\frac{F_{\rm model}(\lambda_i,w_{\rm f},p_{\rm v},f_{\rm r},\tilde{\Gamma})
    -F_{\rm obs}(\lambda_{i})}{\sigma_{\lambda_{i}}}\Big]^{2}~,
\label{chi2}
\end{equation}
to assess the fitting degree of model results with respect to the observations.
The input parameters that gives the minimum $\chi^{2}_{\rm r}$ could be treated
as the most possible values of these parameters.

\subsubsection{Mean Grain radius}
Considering that the surface temperature could differ largely at different
region and thermal inertia is strongly temperature dependent, the obtained mean
thermal inertia from the above radiometry process may not well reveal the physical
condition of the surface materials. Thus it is necessary to remove the temperature
effect so as to find the more basic properties that is not affected by temperature.

As mentioned in above section, if we know the taxonomic type of a small body from
spectral observation, we could find its most spectral-resembled chondrite to
estimate the specific heat capacity $c_{\rm p}(T)$ at various temperature.

Moreover, its surface mass density can also be estimated via
\begin{equation}
\rho=(1-\phi)\rho_{\rm m},
\end{equation}
where the material density $\rho_{\rm m}$ can be approximated as the density
of the corresponding chondrite, such as \citep{Opeil2010}
\[\rho_{\rm m}=3110~{\rm kgm^{-3}}~{\rm for~C-type}, \]
\[\rho_{\rm m}=3700~{\rm kgm^{-3}}~{\rm for~S-type}, \]
and the porosity $\phi$ can have values from 0.4 to 0.6 for the surface
layers of airless bodies.

Finally, if the surface is covered by dust mantle, the thermal conductivity can be
related to the mean grain radius $b$ like Equation (\ref{ktrphi}), meaning that $b$
would be mainly important free parameter that decides the realistic thermal inertia.

Then we can investigate the three free parameters --- mean grain radius, roughness
fraction, and geometric albedo (or effective diameter) in the fitting process.
Still we use the so-called reduced $\chi^{2}_{\rm r}$ defined as
\begin{equation}
\chi^{2}_{\rm r}=\frac{1}{n-3}\sum^{n}_{i=1}
\Big[\frac{F_{\rm model}(\lambda_i,w_{\rm f},p_{\rm v},f_{\rm r},\tilde{b})
    -F_{\rm obs}(\lambda_{i})}{\sigma_{\lambda_{i}}}\Big]^{2}~,
\end{equation}
to assess the fitting degree of model results with respect to the observations.
The input parameters that gives the minimum $\chi^{2}_{\rm r}$ could be treated
as the most possible values of these parameters.

\section{Application example}
As an example, we apply RSTPM to study the main belt object (24) Themis, which has
been believed to be the parent body of most currently known main-belt comets (MBCs),
and hence should have a dust mantle on the surface. MBCs are so small that observations
of them are difficult to obtain, whereas Themis is bright enough to be observed at both
optical and infrared bands. So using thermal infrared observations to study surface dust
properties would be easier for Themis, the results of which can be served as a reference
for the surface dust properties of MBCs regarding their possible connections.

\begin{figure}[htbp]
\includegraphics[scale=0.58]{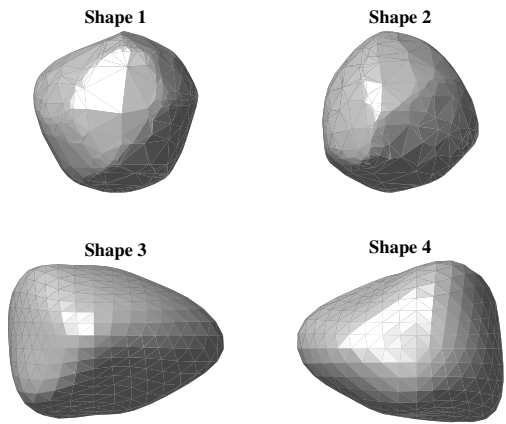}
  \centering
  \caption{The four shape models of (24) Themis from the Database of Asteroid Models from Inversion Techniques. The shape models are shown in the same on-sky orientation.}
  \label{shapefs}
\end{figure}

Although light-curve observations of Themis have been obtained to derive its spin
orientation together with shape model by the light-curve inversion method, there is
still no unique solution to the spin orientation as yet. Currently there exist four
different solutions to the spin orientation and shape model of Themis, as shown in
Table \ref{shapes} and Figure \ref{shapefs}. The lucky thing is that WISE/NEOWISE has
got multi-epoch thermal lightcurves of Themis, thus enabling us to find out which of
the four shape models is the best by using RSTPM to fit these thermal lightcurves.
\begin{table}[htbp]
 \centering
 \renewcommand\arraystretch{0.8}
 \caption{Light-curve inversion shape models of (24) Themis \citep{Hanus2016,Viikinkoski2017}. The shape models can be obtained from the Database of Asteroid Models from Inversion Techniques.}
 \label{shapes}
 \begin{tabular}{@{}lccc@{}}
 \hline
        & Spin Orientation  &  Number  & Number  \\
        & ($\lambda$, $\beta$) ($^\circ$) & of Vertices & of Facets  \\
 \hline
 shape 1  & (331, 52) & 1018 & 511 \\
 shape 2  & (137, 59) & 1018 & 511 \\
 shape 3  & (139, 71) & 800 & 402  \\
 shape 4  & (329, 70) & 800 & 402  \\
 \hline
 \multicolumn{4}{l}{$\lambda$: Ecliptic Longitude;~~$\beta$: Ecliptic Latitude.} \\
 \multicolumn{4}{l}{Spin period: 8.374187 hr} \\
\end{tabular}
\end{table}

\subsection{WISE/NEOWISE observation}
The {\it Wide-field Infrared Survey Explorer} (WISE) mission has mapped entire sky in
four bands at 3.4 (W1), 4.6 (W2), 12 (W2), and 22 (W2) $\mu$m with resolutions from
$6.1^{\prime\prime}$ to $12^{\prime\prime}$. All four bands were imaged simultaneously,
and the exposure times were 7.7 s in 3.4 and 4.6 $\mu$m and 8.8 s in 12 and 22 $\mu$m.
The four-bands survey started from 2010 January 7, and ended on 2010 August 6 after the
outer cryogen tank was exhausted, making the W4 channel be no longer able to be used to
obtain survey data. The W3 channel continued operation until 2010 September 29 when
the inner cryogen reserve was exhausted, while the W1 and W2 channel kept working
until the telescope was set into hibernation on 2011 February 1. The two-band survey
was then resumed on 2013 December 13 (known as NEOWISE), and is still in service, which
has obtained nearly 6-year observations.

We found multi-year observations of Themis from WISE archive (see the website of the
NASA/IPAC Infrared Science Archive http://irsa.ipac.caltech.edu/). The magnitude data are
converted to flux with colour corrections (W1: 2.0577; W2: 1.3448; W3: 1.0006; W4: 0.9833),
and all the derived monochromatic flux densities are set with an associated uncertainty
of $\pm$10 percent \citep{Wright2010}. The flux data are summarized in Table \ref{obs1},
\ref{obs2} and \ref{obs3}.

\begin{table*}[htbp]
\centering
\renewcommand\arraystretch{0.8}
\caption{Mid-infrared observations of (24) Themis: WISE. The epoch marked by red is used as reference epoch to generate thermal light curves.}
\label{obs1}
\begin{tabular}{@{}ccccccccc@{}}
\hline
 UT  & \multicolumn{4}{c}{Flux } & $MA$ & $r_{\rm helio}$ & $\Delta_{\rm obs}$ & $\alpha$  \\
     & 3.4 $\mu$m (mJy)& 4.6 $\mu$m (mJy)& 12 $\mu$m (Jy)& 22 $\mu$m (Jy) & ($^{\circ}$) & (AU) & (AU) & ($^{\circ}$)  \\
\hline
2010-04-21 00:45 & 3.59$\pm$0.36 & 9.81$\pm$0.98 & 4.95$\pm$0.49 & 13.22$\pm$1.32 & 131.584 & 3.421 & 3.278 & 17.082 \\
2010-04-21 13:27 & 3.98$\pm$0.40 & 10.50$\pm$1.05 & 5.02$\pm$0.50 & 15.05$\pm$1.51 & 131.673 & 3.421 & 3.272 & 17.083 \\
2010-04-21 15:02 & 3.65$\pm$0.36 & 10.95$\pm$1.10 & 6.20$\pm$0.62 & 15.99$\pm$1.60 & 131.687 & 3.421 & 3.270 & 17.083 \\
2010-04-21 16:37 & 3.56$\pm$0.36 & 10.57$\pm$1.06 & 6.05$\pm$0.61 & 16.28$\pm$1.63 & 131.702 & 3.421 & 3.269 & 17.083 \\
\textcolor{red}{2010-04-21 18:13} & 3.67$\pm$0.37 & 10.45$\pm$1.05 & 5.29$\pm$0.53 & 15.62$\pm$1.56 & 131.710 & 3.421 & 3.269 & 17.083 \\
2010-04-21 19:48 & 3.69$\pm$0.37 & 11.13$\pm$1.11 & 6.25$\pm$0.62 & 17.08$\pm$1.71 & 131.724 & 3.421 & 3.268 & 17.083 \\
2010-04-21 21:23 & 3.66$\pm$0.37 & 10.25$\pm$1.03 & 5.73$\pm$0.57 & 15.50$\pm$1.55 & 131.732 & 3.421 & 3.267 & 17.083 \\
2010-04-22 00:34 & 3.92$\pm$0.39 & 10.81$\pm$1.08 & 5.99$\pm$0.60 & 16.03$\pm$1.60 & 131.761 & 3.421 & 3.265 & 17.083 \\
2010-04-22 03:44 & 3.98$\pm$0.40 & 12.70$\pm$1.27 & 6.30$\pm$0.63 & 17.28$\pm$1.73 & 131.783 & 3.422 & 3.263 & 17.083 \\
\hline
\multicolumn{9}{c}{$MA$: Orbital Mean Anomaly; $\alpha$: solar phase angle.}
\end{tabular}
\end{table*}
\begin{table}\footnotesize
\centering
\renewcommand\arraystretch{0.8}
\caption{Mid-infrared observations of (24) Themis: NEOWISE (2014-2016).}
\label{obs2}
\begin{tabular*}{8.8cm}{@{}c@{\hspace{0.2cm}}c@{\hspace{0.2cm}}c@{\hspace{0.2cm}}
c@{\hspace{0.2cm}}c@{\hspace{0.2cm}}c@{\hspace{0.2cm}}c@{}}
\hline
 UT  & \multicolumn{2}{c}{Flux (mJy)} & $MA$ & $r_{\rm helio}$ & $\Delta_{\rm obs}$ & $\alpha$  \\
     & 3.4 $\mu$m & 4.6 $\mu$m & ($^{\circ}$) & (AU) & (AU) & ($^{\circ}$)  \\
\hline
2014-06-10 20:42 & 7.65$\pm$0.76 & 38.43$\pm$3.84 & 39.354 & 2.854 & 2.633 & -20.826 \\
2014-06-10 20:43 & 7.80$\pm$0.78 & 40.10$\pm$4.01 & 39.354 & 2.854 & 2.633 & -20.826 \\
2014-06-10 23:52 & 7.48$\pm$0.75 & 36.84$\pm$3.68 & 39.376 & 2.854 & 2.635 & -20.826 \\
2014-06-11 03:01 & 7.47$\pm$0.75 & 36.10$\pm$3.61 & 39.398 & 2.854 & 2.636 & -20.827 \\
2014-06-11 06:11 & 7.54$\pm$0.75 & 41.18$\pm$4.12 & 39.421 & 2.854 & 2.638 & -20.828 \\
2014-06-11 07:46 & 7.25$\pm$0.73 & 34.22$\pm$3.42 & 39.435 & 2.854 & 2.639 & -20.828 \\
2014-06-11 09:20 & 7.21$\pm$0.72 & 36.07$\pm$3.61 & 39.443 & 2.854 & 2.640 & -20.829 \\
2014-06-11 10:55 & 7.39$\pm$0.74 & 35.15$\pm$3.51 & 39.458 & 2.854 & 2.641 & -20.828 \\
\textcolor{red}{2014-06-11 12:30} & 7.60$\pm$0.76 & 37.15$\pm$3.71 & 39.472 & 2.855 & 2.642 & -20.828 \\
2014-06-11 14:05 & 7.48$\pm$0.75 & 41.30$\pm$4.13 & 39.480 & 2.855 & 2.643 & -20.829 \\
2014-06-11 15:39 & 7.25$\pm$0.73 & 35.08$\pm$3.51 & 39.495 & 2.855 & 2.644 & -20.828 \\
2014-06-11 17:14 & 7.53$\pm$0.75 & 37.28$\pm$3.73 & 39.502 & 2.855 & 2.644 & -20.829 \\
2014-06-11 20:24 & 7.21$\pm$0.72 & 35.08$\pm$3.51 & 39.524 & 2.855 & 2.646 & -20.830 \\
2014-06-11 23:33 & 7.63$\pm$0.76 & 37.28$\pm$3.73 & 39.554 & 2.855 & 2.649 & -20.829 \\
2014-06-11 23:33 & 7.63$\pm$0.76 & 37.28$\pm$3.73 & 39.554 & 2.855 & 2.649 & -20.829 \\
2014-06-12 02:43 & 7.48$\pm$0.75 & 36.07$\pm$3.61 & 39.576 & 2.855 & 2.650 & -20.829 \\
\\
2015-08-31 14:31 & 4.48$\pm$0.45 & 12.51$\pm$1.25 & 118.614 & 3.360 & 3.066 & -17.304 \\
2015-09-01 01:32 & 4.83$\pm$0.48 & 11.53$\pm$1.15 & 118.695 & 3.360 & 3.072 & -17.316 \\
2015-09-01 03:07 & 4.23$\pm$0.42 & 11.02$\pm$1.10 & 118.702 & 3.360 & 3.073 & -17.318 \\
2015-09-01 06:16 & 4.05$\pm$0.40 & 11.05$\pm$1.11 & 118.725 & 3.360 & 3.075 & -17.321 \\
\textcolor{red}{2015-09-01 06:16} & 4.05$\pm$0.40 & 11.05$\pm$1.11 & 118.725 & 3.360 & 3.075 & -17.321 \\
2015-09-01 07:50 & 3.21$\pm$0.32 & 11.34$\pm$1.13 & 118.739 & 3.361 & 3.076 & -17.323 \\
2015-09-01 10:59 & 4.18$\pm$0.42 & 10.98$\pm$1.10 & 118.762 & 3.361 & 3.078 & -17.326 \\
2015-09-01 14:08 & 4.89$\pm$0.49 & 11.17$\pm$1.12 & 118.784 & 3.361 & 3.080 & -17.329 \\
2015-09-01 17:17 & 4.05$\pm$0.40 & 9.67$\pm$0.97 & 118.806 & 3.361 & 3.082 & -17.332 \\
2015-09-01 17:17 & 4.05$\pm$0.40 & 9.67$\pm$0.97 & 118.806 & 3.361 & 3.082 & -17.332 \\
\\
2016-05-17 11:02 & 3.24$\pm$0.32 & 7.95$\pm$0.79 & 164.647 & 3.519 & 3.384 & 16.700 \\
2016-05-17 14:10 & 3.45$\pm$0.34 & 8.73$\pm$0.87 & 164.669 & 3.519 & 3.383 & 16.701 \\
2016-05-17 17:19 & 3.39$\pm$0.34 & 7.93$\pm$0.79 & 164.691 & 3.519 & 3.381 & 16.701 \\
2016-05-17 17:19 & 3.39$\pm$0.34 & 7.93$\pm$0.79 & 164.691 & 3.519 & 3.381 & 16.701 \\
2016-05-17 18:53 & 3.26$\pm$0.33 & 8.01$\pm$0.80 & 164.706 & 3.519 & 3.380 & 16.702 \\
\textcolor{red}{2016-05-17 20:28} & 3.22$\pm$0.32 & 7.62$\pm$0.76 & 164.713 & 3.519 & 3.379 & 16.702 \\
2016-05-17 22:02 & 3.34$\pm$0.33 & 8.42$\pm$0.84 & 164.728 & 3.519 & 3.378 & 16.702 \\
2016-05-17 23:37 & 3.43$\pm$0.34 & 7.78$\pm$0.78 & 164.743 & 3.519 & 3.377 & 16.703 \\
2016-05-18 01:11 & 3.31$\pm$0.33 & 7.66$\pm$0.77 & 164.750 & 3.519 & 3.376 & 16.702 \\
2016-05-18 04:19 & 3.18$\pm$0.32 & 8.02$\pm$0.80 & 164.772 & 3.519 & 3.374 & 16.703 \\
2016-05-18 07:28 & 3.47$\pm$0.35 & 8.65$\pm$0.86 & 164.794 & 3.520 & 3.373 & 16.703 \\
\\
2016-10-26 01:06 & 3.87$\pm$0.39 & 8.87$\pm$0.89 & -166.726 & 3.522 & 3.137 & -15.848 \\
2016-10-26 04:15 & 4.00$\pm$0.40 & 9.37$\pm$0.94 & -166.704 & 3.522 & 3.138 & -15.856 \\
2016-10-26 07:24 & 3.93$\pm$0.39 & 9.06$\pm$0.91 & -166.682 & 3.522 & 3.140 & -15.864 \\
2016-10-26 08:58 & 3.57$\pm$0.36 & 8.51$\pm$0.85 & -166.667 & 3.522 & 3.141 & -15.869 \\
\textcolor{red}{2016-10-26 10:32} & 3.59$\pm$0.36 & 8.07$\pm$0.81 & -166.653 & 3.522 & 3.143 & -15.875 \\
2016-10-26 12:07 & 3.98$\pm$0.40 & 10.18$\pm$1.02 & -166.645 & 3.522 & 3.143 & -15.877 \\
2016-10-26 13:41 & 3.70$\pm$0.37 & 7.85$\pm$0.78 & -166.630 & 3.522 & 3.144 & -15.883 \\
2016-10-26 15:15 & 4.11$\pm$0.41 & 8.78$\pm$0.88 & -166.623 & 3.522 & 3.145 & -15.885 \\
2016-10-26 18:24 & 3.64$\pm$0.36 & 8.27$\pm$0.83 & -166.601 & 3.522 & 3.147 & -15.893 \\
2016-10-27 00:41 & 3.75$\pm$0.38 & 8.76$\pm$0.88 & -166.549 & 3.522 & 3.151 & -15.911 \\
\hline
\multicolumn{7}{c}{$MA$: Orbital Mean Anomaly; $\alpha$: solar phase angle.}
\end{tabular*}
\end{table}

\begin{table}\footnotesize
\centering
\renewcommand\arraystretch{0.8}
\caption{Mid-infrared observations of (24) Themis: NEOWISE (2017-2018).}
\label{obs3}
\begin{tabular*}{8.6cm}{@{}c@{\hspace{0.2cm}}c@{\hspace{0.2cm}}c@{\hspace{0.2cm}}
c@{\hspace{0.2cm}}c@{\hspace{0.2cm}}c@{\hspace{0.2cm}}c@{}}
\hline
 UT  & \multicolumn{2}{c}{Flux (mJy)} & $MA$ & $r_{\rm helio}$ & $\Delta_{\rm obs}$ & $\alpha$  \\
     & 3.4 $\mu$m & 4.6 $\mu$m & ($^{\circ}$) & (AU) & (AU) & ($^{\circ}$)  \\
\hline
2017-07-23 04:59 & 3.78$\pm$0.38 & 10.48$\pm$1.05 & -118.84 & 3.361 & 3.199 & 17.596 \\
2017-07-23 08:07 & 3.70$\pm$0.37 & 10.44$\pm$1.04 & -118.82 & 3.360 & 3.197 & 17.596 \\
2017-07-23 11:16 & 3.84$\pm$0.38 & 11.18$\pm$1.12 & -118.79 & 3.360 & 3.195 & 17.596 \\
2017-07-23 14:24 & 3.99$\pm$0.40 & 11.52$\pm$1.15 & -118.77 & 3.360 & 3.194 & 17.597 \\
2017-07-23 17:33 & 3.79$\pm$0.38 & 10.50$\pm$1.05 & -118.74 & 3.360 & 3.191 & 17.597 \\
2017-07-23 19:07 & 3.78$\pm$0.38 & 11.04$\pm$1.10 & -118.74 & 3.360 & 3.190 & 17.597 \\
\textcolor{red}{2017-07-23 20:41} & 3.93$\pm$0.39 & 10.80$\pm$1.08 & -118.72 & 3.360 & 3.189 & 17.597 \\
2017-07-23 22:15 & 3.76$\pm$0.38 & 10.83$\pm$1.08 & -118.72 & 3.360 & 3.188 & 17.597 \\
2017-07-24 01:24 & 3.74$\pm$0.37 & 10.08$\pm$1.01 & -118.69 & 3.360 & 3.187 & 17.597 \\
2017-07-24 04:33 & 3.79$\pm$0.38 & 10.69$\pm$1.07 & -118.66 & 3.360 & 3.184 & 17.597 \\
2017-07-24 07:41 & 4.06$\pm$0.41 & 12.58$\pm$1.26 & -118.64 & 3.360 & 3.182 & 17.597 \\
2017-07-24 10:50 & 4.08$\pm$0.41 & 11.20$\pm$1.12 & -118.62 & 3.359 & 3.180 & 17.596 \\
\\
2017-12-22 12:18 & 5.93$\pm$0.59 & 17.34$\pm$1.73 & -91.832 & 3.198 & 2.731 & -16.854 \\
2017-12-22 15:27 & 6.02$\pm$0.60 & 18.06$\pm$1.81 & -91.809 & 3.197 & 2.733 & -16.867 \\
2017-12-22 18:35 & 6.36$\pm$0.64 & 21.22$\pm$2.12 & -91.780 & 3.197 & 2.735 & -16.885 \\
2017-12-22 20:10 & 5.80$\pm$0.58 & 16.43$\pm$1.64 & -91.772 & 3.197 & 2.735 & -16.890 \\
2017-12-22 21:44 & 6.03$\pm$0.60 & 18.28$\pm$1.83 & -91.758 & 3.197 & 2.736 & -16.899 \\
2017-12-22 23:18 & 6.01$\pm$0.60 & 18.29$\pm$1.83 & -91.750 & 3.197 & 2.737 & -16.903 \\
2017-12-23 00:52 & 5.80$\pm$0.58 & 17.06$\pm$1.71 & -91.735 & 3.197 & 2.738 & -16.912 \\
\textcolor{red}{2017-12-23 00:53} & 5.75$\pm$0.58 & 17.57$\pm$1.76 & -91.735 & 3.197 & 2.738 & -16.912 \\
2017-12-23 02:27 & 6.15$\pm$0.62 & 20.72$\pm$2.07 & -91.728 & 3.197 & 2.739 & -16.916 \\
2017-12-23 02:27 & 6.15$\pm$0.62 & 20.72$\pm$2.07 & -91.728 & 3.197 & 2.739 & -16.916 \\
2017-12-23 05:35 & 6.16$\pm$0.62 & 18.08$\pm$1.81 & -91.698 & 3.197 & 2.741 & -16.934 \\
2017-12-23 08:44 & 5.77$\pm$0.58 & 17.50$\pm$1.75 & -91.676 & 3.196 & 2.742 & -16.947 \\
2017-12-23 11:52 & 5.93$\pm$0.59 & 19.16$\pm$1.92 & -91.654 & 3.196 & 2.744 & -16.960 \\
\\
2018-10-17 10:08 & 7.00$\pm$0.70 & 39.33$\pm$3.93 & -38.793 & 2.854 & 2.685 & 20.440 \\
2018-10-17 13:17 & 7.17$\pm$0.72 & 41.76$\pm$4.18 & -38.770 & 2.853 & 2.684 & 20.441 \\
2018-10-17 16:25 & 7.42$\pm$0.74 & 39.55$\pm$3.95 & -38.748 & 2.853 & 2.682 & 20.441 \\
2018-10-17 19:34 & 8.19$\pm$0.82 & 42.65$\pm$4.27 & -38.719 & 2.853 & 2.680 & 20.441 \\
2018-10-17 19:34 & 8.19$\pm$0.82 & 42.65$\pm$4.27 & -38.719 & 2.853 & 2.680 & 20.441 \\
2018-10-17 22:42 & 7.28$\pm$0.73 & 38.26$\pm$3.83 & -38.697 & 2.853 & 2.678 & 20.441 \\
2018-10-18 00:16 & 7.20$\pm$0.72 & 40.73$\pm$4.07 & -38.689 & 2.853 & 2.677 & 20.442 \\
\textcolor{red}{2018-10-18 01:51} & 7.43$\pm$0.74 & 40.02$\pm$4.00 & -38.674 & 2.853 & 2.676 & 20.441 \\
2018-10-18 03:24 & 7.44$\pm$0.74 & 40.99$\pm$4.10 & -38.667 & 2.853 & 2.675 & 20.442 \\
2018-10-18 04:58 & 7.48$\pm$0.75 & 46.38$\pm$4.64 & -38.652 & 2.853 & 2.674 & 20.442 \\
2018-10-18 06:33 & 7.29$\pm$0.73 & 40.06$\pm$4.01 & -38.637 & 2.853 & 2.673 & 20.442 \\
2018-10-18 09:42 & 7.38$\pm$0.74 & 39.37$\pm$3.94 & -38.615 & 2.853 & 2.671 & 20.442 \\
2018-10-18 12:50 & 8.44$\pm$0.84 & 44.46$\pm$4.45 & -38.593 & 2.853 & 2.669 & 20.442 \\
2018-10-18 15:58 & 7.66$\pm$0.77 & 41.68$\pm$4.17 & -38.571 & 2.852 & 2.667 & 20.442 \\
\hline
\multicolumn{7}{c}{$MA$: Orbital Mean Anomaly; $\alpha$: solar phase angle.}
\end{tabular*}
\end{table}

\subsection{Input parameters}
In order to interpret these multi-year observations, RSTPM needs several input
parameters, including observation geometry, shape model, spin orientation, rotation
phase $ph$, scattering weight-factor $w_{\rm f}$, geometric albedo $p_{\rm v}$, roughness
fraction $f_{\rm r}$, and mean grain radius $\tilde{b}$.

The observation geometry at the time of each observation can be easily obtained according
to the orbit of Themis and WISE. Spin orientation together with shape model has four different
choices as listed in Table \ref{shapes}, which makes the rotation phase of each observation
unclear as well. Hence the spin orientation would be the first parameter that need to be
investigated by the fitting procedure.

The scattering weight-factor $w_{\rm f}$ is crucial in fitting procedure, as the W1 and W2 band
observations contain significant amount of sunlight reflection. Since this parameter is
an artificial factor used to interpret sunlight refection, its physical significance isn't
that clear, thus we only need a scattering weight-factor $w_{\rm f}$ that could achieve best-fitting
degree to the observations.

The other parameters including geometric albedo $p_{\rm v}$, roughness fraction $f_{\rm r}$,
and mean grain radius $\tilde{b}$, are all free parameters that would be determined by
optimization of fitting process.

\subsection{Fitting with rotationally averaged flux: seasonal effect}
As mentioned above, rotation phases at different observation epochs are unclear due to the
uncertainties of spin orientation, thus at the fist step, we choose the rotationally averaged
model flux $F_{\rm model}$ to fit the observations, by which the diurnal effect is eliminated,
whereas the seasonal effect is highlighted to investigate the influence of spin orientation.

\subsubsection{Best-fit spin orientation}
\begin{figure}[htbp]
\includegraphics[scale=0.58]{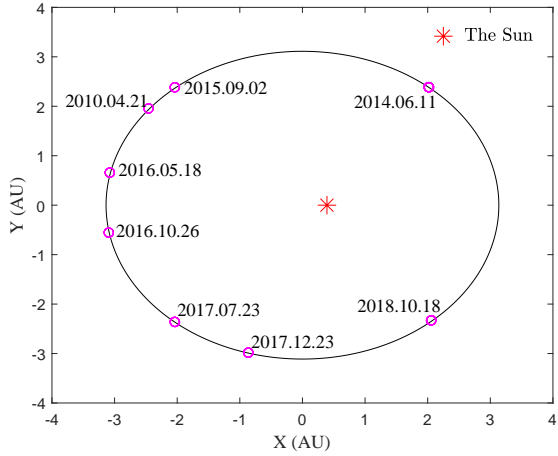}
  \centering
  \caption{The little magenta circles represent orbital positions of (24) Themis at the time of each observation, covering 8 different epochs, making it possible to resolve the spin orientation, roughness fraction and thermal parameters by considering the seasonal variation of observed infrared flux by WISE/NEOWISE.}
  \label{obsOrbPos}
\end{figure}

The available WISE/NEOWISE observations of Themis cover nearly 8 different epochs, as
shown in Figure \ref{obsOrbPos}. Therefore, these observations of infrared flux can show
seasonal variation, making it possible for us to investigate the probable spin orientation,
roughness fraction and thermal parameters.

By fitting observations with rotationally averaged model flux generated by RSTPM
under input of different spin orientation and physical parameters, best-fit results are
selected out and summarized in Table \ref{minirchi2}.
\begin{table}[htbp]
 \centering
 \renewcommand\arraystretch{0.8}
 \caption{Best-fit results by fitting observations with rotationally averaged model flux.}
 \label{minirchi2}
 \begin{tabular}{@{}lccccc@{}}
 \hline
        & \multicolumn{4}{c}{Best-fitting parameters}  &  Minimum  \\
        & $w_{\rm f}$ & $p_{\rm v}$ & $f_{\rm r}$ & $\tilde{b}$ ($\mu$m)& $\chi^2_{\rm r}$\\
 \hline
 shape 1  & 0.37 & 0.067 & 0.45 & 150 &  0.383 \\
 shape 2  & 0.32 & 0.064 & 0.40 & 140 &  0.291 \\
 shape 3  & 0.50 & 0.058 & 0.50 & 140 &  0.982  \\
 shape 4  & 0.50 & 0.060 & 0.50 & 140 &  0.842  \\
 \hline
 \multicolumn{6}{l}{$w_{\rm f}$: scattering weight-factor.} \\
 \multicolumn{6}{l}{$f_{\rm r}$: roughness fraction;~~$\tilde{b}$: mean grain radius.} \\
\end{tabular}
\end{table}

According to Table \ref{minirchi2}, in the case of shape 2, minimum reduced $\chi^2_{\rm r}$
is much smaller than that obtained in other three cases, indicating that shape 2 with spin
orientation ($\lambda=137^\circ$, $\beta=59^\circ$) should be the best solution to the spin
orientation and shape model of Themis.
This result is consistent with the recent work of \citet{ORourke2020}, which also
concluded ($\lambda=137^\circ$, $\beta=59^\circ$) to be the best-fit spin orientation by
TPM fitting to the Subaru/COMICS observations of Themis, indicating that this best solution
of spin orientation is not accidental.

On the other hand, for the case of shape 2, we obtain best-fit scattering weight-factor
$w_{\rm f}=0.32$. Then, in the following sections, we will utilize shape 2 and $w_{\rm f}=0.32$
to further study the geometric albedo $p_{\rm v}$, roughness fraction $f_{\rm r}$, mean grain
radius $\tilde{b}$ and thermal inertia $\Gamma$ of Themis.

\subsubsection{Results of $p_{\rm v}$,$f_{\rm r}$, $\tilde{b}$}
\begin{figure}[htbp]
\includegraphics[scale=0.58]{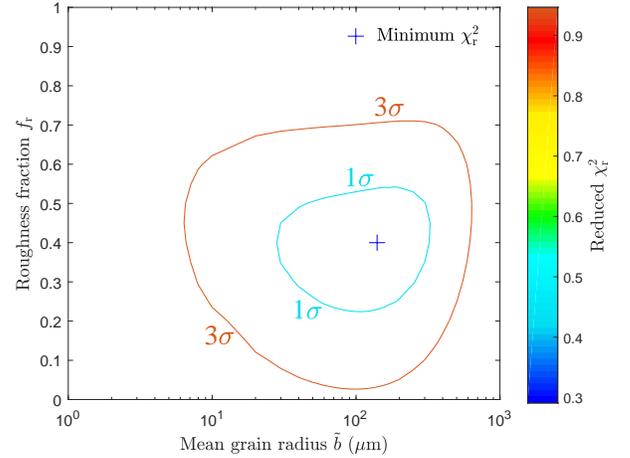}
  \centering
  \caption{The contour of $\chi^2_{\rm r}$($f_{\rm r}$,$\tilde{b}$), which is obtained by fitting the observations with two free parameters: roughness fraction $f_{\rm r}$ and mean grain radius $\tilde{b}$. The cyan region stands for the $1\sigma$-level constraint, and the red region represents the $3\sigma$-level constraint.}
  \label{contourFrb}
\end{figure}

By fixing shape model with spin orientation ($\lambda=137^\circ$, $\beta=59^\circ$) and
scattering weight-factor $w_{\rm f}=0.32$, we then fit the observations by scanning roughness
fraction $f_{\rm r}$ in the range of $0\sim1$ and mean grain radius $\tilde{b}$ in the
range of $1\sim1000~\mu$m. With each pair of ($f_{\rm r}$,$\tilde{b}$), a best-fit
geometric albedo $p_{\rm v}$ together with effective diameter $D_{\rm eff}$ is found to
compute the reduced $\chi^2_{\rm r}$. The results are presented in Figure \ref{contourFrb}
as a contour of $\chi^2_{\rm r}$($f_{\rm r}$,$\tilde{b}$).

According to Figure \ref{contourFrb}, a well constrained $1\sigma$-level limit is
derived for roughness fraction $f_{\rm r}$ and mean grain radius $\tilde{b}$, giving
$f_{\rm r}=0.4\pm0.15$ (corresponding to RMS slope $31\pm6$),
$\tilde{b}=140^{+180}_{-110}~\mu$m respectively.
$3\sigma$-level constraint for roughness fraction is derived as
$f_{\rm r}=0.4^{+0.3}_{-0.4}$ (corresponding to RMS slope $31^{+11}_{-31}$),
whereas for mean grain radius, a relatively wide $3\sigma$-level limit is obtained as
$\tilde{b}=140^{+500}_{-114}(6\sim640)~\mu$m.

\begin{figure}[htbp]
\includegraphics[scale=0.58]{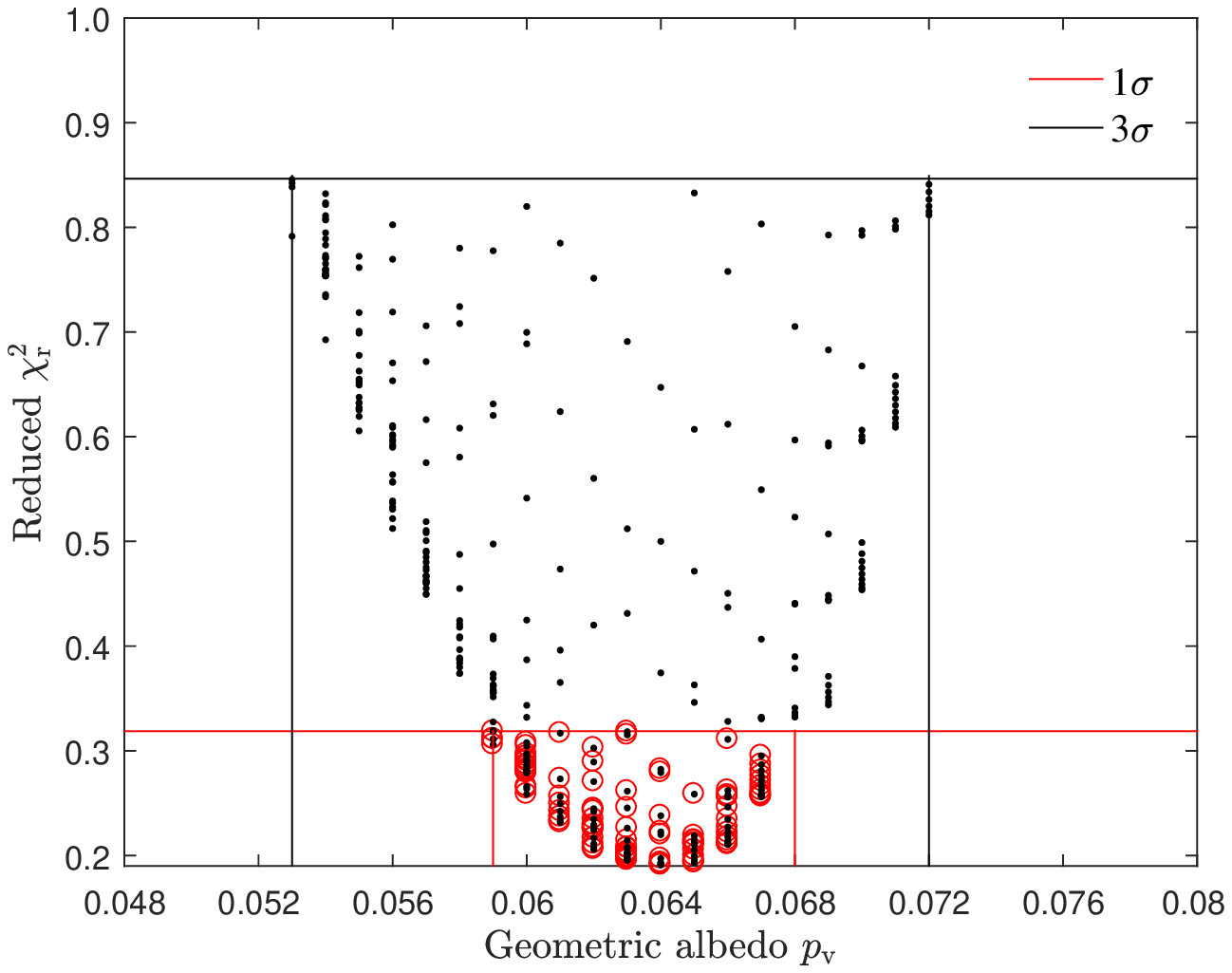}
  \centering
  \caption{$p_{\rm v}\sim\chi^2_{\rm reduced}$ profiles fit to the observations in consideration of the derived $1\sigma$ and $3\sigma$ ranges of roughness fraction and mean grain radius.}
  \label{galbedo}
\end{figure}

According to the above derived $1\sigma$ and $3\sigma$ ranges of roughness fraction
and mean grain radius, the corresponding geometric albedo $p_{\rm v}$ and $\chi^2_{\rm r}$
are picked out, leading to the $p_{\rm v}\sim\chi^2_{\rm r}$ relation as shown in
Figure \ref{galbedo}. In this way, we obtain the $1\sigma$ and $3\sigma$-level limits
of geometric albedo as $p_{\rm v}=0.064^{+0.004}_{-0.005}$ and
$p_{\rm v}=0.064^{+0.008}_{-0.011}$ respectively, and consequently the effective
diameter of Themis can be derived to be $D_{\rm eff}=201.6^{+8.4}_{-5.9}$ km ($1\sigma$) and
$D_{\rm eff}=201.6^{+19.9}_{-11.5}$ km ($3\sigma$) in consideration of the absolute visual
magnitude $H_{\rm v}=7.08$ and slope parameter $G=0.19$ \citep{Harris1989}.

Our result of geometric albedo $p_{\rm v}=0.064^{+0.008}_{-0.011}$ of Themis agrees with
the result $p_{\rm v}=0.07\pm0.01$ of \citet{ORourke2020} in the range $0.06\sim0.072$,
despite that our result tends to be smaller, which consequently leads to a larger estimation of
effective diameter $D_{\rm eff}=201.6^{+19.9}_{-11.5}$ km in comparison to the result
$D_{\rm eff}=192^{+10}_{-7}$ km derived by \citet{ORourke2020}. Differences including data inputs
and modeling procedures between our work and \citet{ORourke2020} may both contribute to the slight
differences of geometric albedo and effective diameter.

\begin{figure}[htbp]
\includegraphics[scale=0.58]{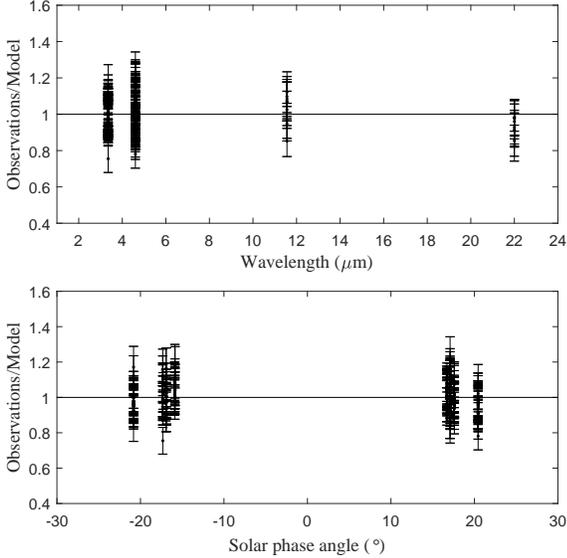}
  \centering
  \caption{The observation/model ratios as a function of wavelength (upper panel) and  solar phase angle (under panel) for the case of best-fit parameters.
  }\label{nsppha}
\end{figure}

To verify the reliability of outcomes derived by the above fitting procedure,
we employ the ratio of observation/model to examine how the model results
match the observations at various observation wavelengths and geometries
(Figure \ref{nsppha}), because these factors are the basic variables of the
observations.

The upper panel of Figure \ref{nsppha} show the observation/model ratios at
each observation wavelength, where the ratios are evenly distributed around
1.0 without significant wavelength dependent features, indicating that
the surface emissivity or albedo of Themis do not show significant wavelength
dependence, thus the combination model of surface thermal emission and sunlight
reflection is good enough to interpret WISE/NEOWISE observations of Themis.

The lower panel of Figure \ref{nsppha} presents the observation/model ratios
at different solar phase angles, where the ratios are also uniformly distributed
round 1.0, showing no distinct phase-angle dependent features, indicating that
the thermal infrared beaming effect of Themis is well resolved by our model,
and hence the expectation of removing the degeneracy between thermal parameters
and roughness by multi-epoch data is well realised. Therefore, it should be
safe to claim that the above fitting procedure and derived results are reliable.

\subsubsection{Seasonal variation of thermal inertia}
As illustrated in section 3.3, thermal inertia is a strong function of temperature.
Now with the above derived profile of mean grain radius, we can evaluate the change
of surface thermal inertia of Themis due to the influence of seasonal temperature variation.

\begin{figure}[htbp]
\includegraphics[scale=0.58]{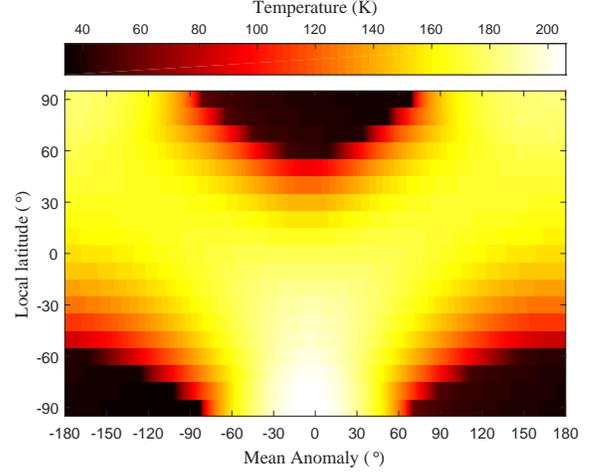}
  \centering
  \caption{Seasonal variation of the diurnal-averaged surface temperature as a function of local latitude. The so-called local latitude is defined as the the complementary angle of the angle between the local normal vector and the rotation axis.}
  \label{Thsst}
\end{figure}

In Figure \ref{Thsst}, a map of surface temperature of Themis is plotted as a function of local
latitude and orbital mean anomaly. Each temperature has been averaged over one rotational
period. We can clearly see that temperature on each local latitude can reach maximum (summer)
or minimum (winter) at different orbital positions as a result of seasonal effect. Temperature
on the poles can vary from $\sim$34 K to $\sim$206 K.

\begin{figure}[htbp]
\includegraphics[scale=0.58]{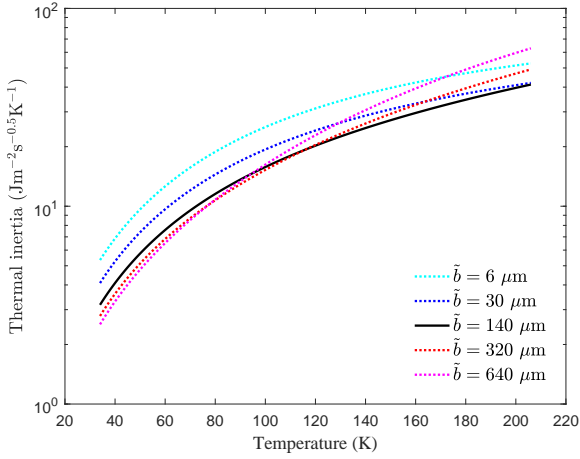}
  \centering
  \caption{Change of surface thermal inertia of Themis due to seasonal temperature variation, considering $3\sigma$-level uncertainties of mean grain radius $\tilde{b}$. }
  \label{GammaSv}
\end{figure}

The significant temperature variation caused by seasonal effect can have influence on
the thermal inertia of surface materials. With Equation (\ref{defgma}), (\ref{ktrphi})
and (\ref{cpt}), variation of surface thermal inertia can be evaluated, as shown in
Figure \ref{GammaSv}. If considering $3\sigma$-level results of mean grain radius
$\tilde{b}$, surface thermal inertia of Themis may vary from a minimum profile of
$\sim3\rm~Jm^{-2}s^{-0.5}K^{-1}$ to a maximum value of $\sim60\rm~Jm^{-2}s^{-0.5}K^{-1}$.
Moreover, Figure \ref{GammaSv} shows that, in comparison to the uncertainties of mean
grain radius, influence of temperature variation on thermal inertia is more significant.

\subsubsection{Average thermal inertia}
\begin{figure}[htbp]
\includegraphics[scale=0.58]{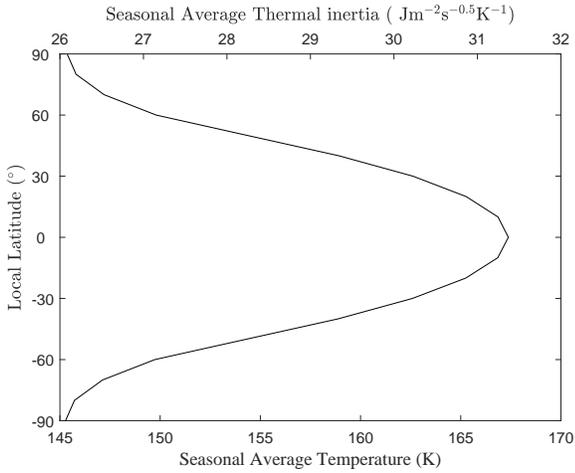}
  \centering
  \caption{Seasonal average temperature and thermal inertia of (24) Themis. }
  \label{ATATh}
\end{figure}

Despite the fact that thermal inertia is temperature dependent, the majority of relevant works
ignore such temperature dependence and only estimate the average thermal inertias.
For comparison with such existing results, we estimate the seasonal average thermal
inertia of Themis by using Equation (\ref{defgma}), (\ref{ktrphi}) and (\ref{cpt}) with
inputs of the derived mean grain radius and seasonal averaged temperature of Themis.
The seasonal average temperature would be a function of local latitude, $\tilde{T}(\theta)$,
and can be estimated as
\begin{equation}
(1-A_{\rm eff,B})\tilde{L}_{\rm s}(\theta)=\varepsilon\sigma \tilde{T}(\theta)^4,
\end{equation}
where $A_{\rm eff,B}$ is the bond albedo, $\varepsilon\sim0.9$ is the average thermal
emissivity, $\tilde{L}_{\rm s}(\theta)$ is the annual average incoming solar flux on
each latitude. The results are presented in Figure \ref{ATATh}.

From Figure \ref{ATATh}, we see that the  average thermal inertia of Themis would
be within $26\sim32\rm~Jm^{-2}s^{-0.5}K^{-1}$. As a comparison, our result of average
thermal inertia is well consistent with that of \citet{ORourke2020}, which estimated
Themis to have a mean thermal inertia of $\sim20^{+25}_{-10}\rm~Jm^{-2}s^{-0.5}K^{-1}$.

\subsection{Fitting with thermal light curve: diurnal effect}
Since the WISE/NEOWISE data at different epochs do not perfectly cover an entire rotation
period and have been observed at various solar phase angles, it is not proper to directly
use them to generate thermal light curves. However, orbital and rotational parameters
of Themis is well known. Thus, in principle, we could derive the rotational phase
of each observation data with respect to a defined local body-fixed coordinate system
if we know the observed rotational phase at a particular epoch. These data could then
be used to create thermal light curves.

\begin{figure}[htbp]
\includegraphics[scale=0.58]{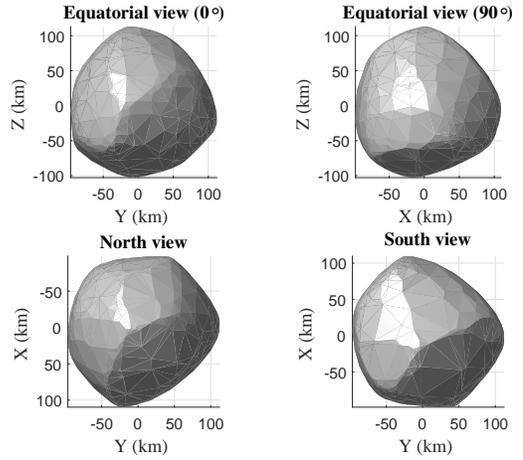}
  \centering
  \caption{3D shape model used to define the local body-fixed coordinate system, where the "zero" rotational phase is chosen to be the Equatorial view $0^\circ$ (view along the x-axis, with the y-axis extending horizontally and the z-axis extending vertically).}
  \label{bodyfixedcs}
\end{figure}

The 3D shape model is used to to define the local body-fixed coordinate system, where
the z-axis is chosen to be the rotation axis, and "zero" rotational phase is chosen
to be the "Equatorial view ($0^\circ$)"  as shown in Figure \ref{bodyfixedcs}.
Moreover, if we define the view angle of one observation with respect to the body-fixed
coordinate system to be $(\varphi,\theta)$, where $\varphi$ stands for local longitude,
and $\theta$ means local latitude, then the rotational phase $ph$ of this observation
can be related to the local longitude $\varphi$ via
\begin{equation}
ph=1-\varphi/(2\pi).
\end{equation}

\begin{figure*}[htbp]
\includegraphics[scale=0.58]{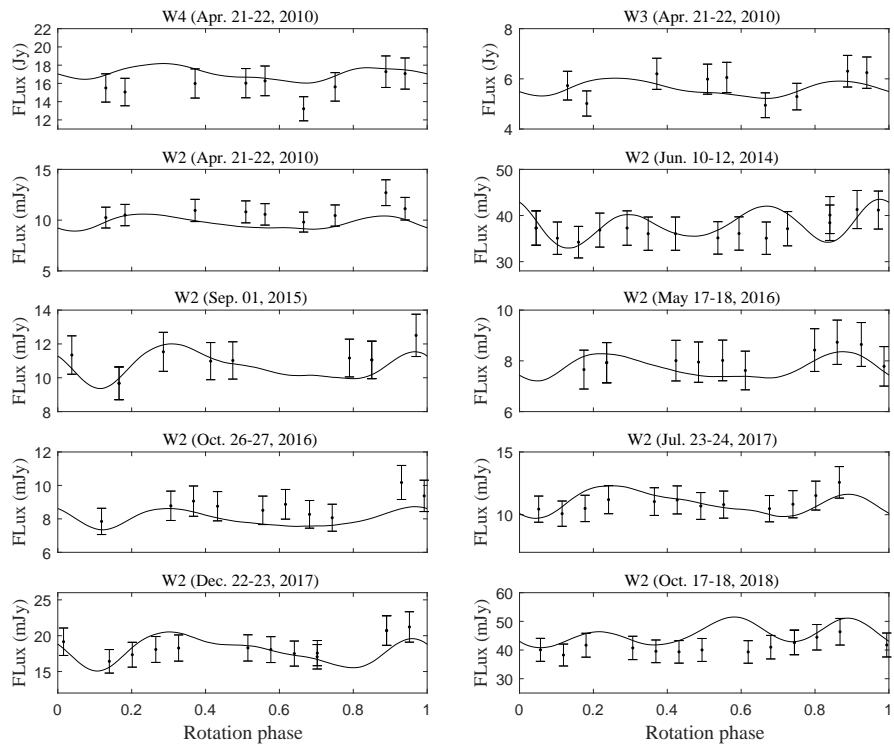}
  \centering
  \caption{Best-fit results to the thermal light curves of WISE/NEOWISE at band W4, W3, W2.
  }\label{thlcsW432}
\end{figure*}

\begin{figure*}[hbp]
\includegraphics[scale=0.58]{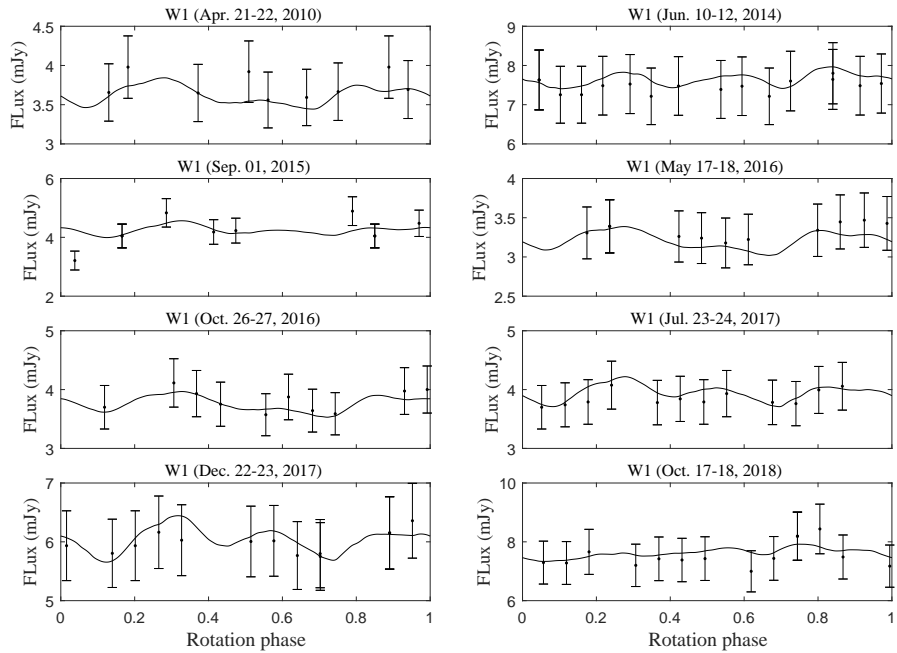}
  \centering
  \caption{Best-fit results to the light curves of WISE/NEOWISE at band W1.
  }\label{tlcsW1}
\end{figure*}

If selecting a reference epoch, and assuming the rotational phase at this epoch to be
$zph$, then all the rotational phases of other data could be derived in consideration
of the observation time and geometry. Furthermore, for some particular epoch, thermal
light curves can be derived for each band by correcting the observed flux at various
epochs into one rotation period at this epoch, where the correction is implemented via
\begin{equation}
F_{i,\rm corr}=F_i\left(\frac{r_{i,\rm helio}}{r_{0,\rm helio}}\right)^2
\left(\frac{\Delta_{i,\rm obs}}{\Delta_{0,\rm obs}}\right)^2,
\label{fluxcorr}
\end{equation}
in which $F_{i,\rm corr}$ is the flux after correction, $F_i$ is the original observed flux,
$r_{i,\rm helio}$ and $r_{0,\rm helio}$ are the heliocentric distance of epoch $i$ and the
reference epoch, while $\Delta_{i,\rm obs}$ and $\Delta_{0,\rm obs}$ are the observation distance.

Following the above method, firstly we select '2014-06-10 20:42' as the reference epoch for
deriving the reference rotational phase $zph$. But for correction of flux, in order to reduce
flux errors caused by correction Equation (\ref{fluxcorr}), we select 8 separate reference epochs,
so as to use data close to each reference epoch (data within three days) to generate thermal
light curves. These 8 reference epochs are marked with red color in Table \ref{obs1}, \ref{obs2}
and \ref{obs3}. Then for each of the reference epoch, theoretical thermal light curves are
simulated by RSTPM to fit the above generated observations of thermal light curves.
The best-fit results are plotted in Figure \ref{thlcsW432} and \ref{tlcsW1}.

Now questions arise what we can learn by fitting with thermal light curves.
Since both irregular shape and surface heterogeneity can contribute to the
rotational variation of flux in light-curves, so we can evaluate whether
the surface has heterogeneity along longitude by fitting with thermal light curves.
To realize this purpose, we can investigate whether the ratios of observation/model
have rotation phase dependent features.

\begin{figure}[htbp]
\includegraphics[scale=0.58]{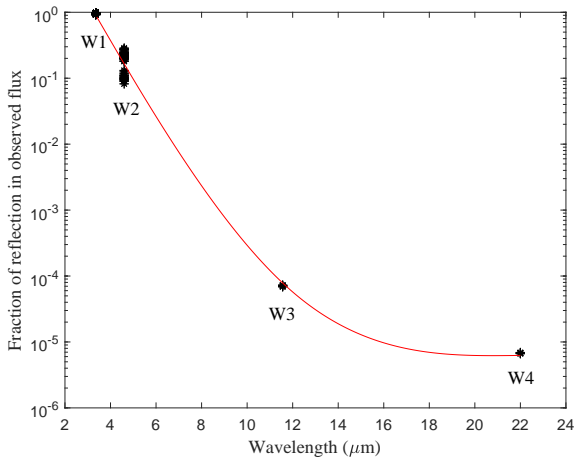}
  \centering
  \caption{The fraction of sunlight-reflection in the observed flux of Themis for each band of WISE/NEOWISE.
  }\label{rlfraction}
\end{figure}

Note that the fraction of sunlight-reflection in each band observation of WISE/NEOWISE
is significantly different. To show such differences, best-fit parameters are input to
RSTPM to estimate the fraction of sunlight-reflection in each band observation
at each epoch. The results are shown in  in Figure \ref{rlfraction}, where
we can see that the fraction of reflection is almost zero at bands W4 and W3 ($<10^{-4}$),
but comes to be non-negligible at band W2 ($\sim10-30\%$), and becomes dominating at band
W1 (reaches up to $\sim99\%$), so the observation/model ratios for bandsW4 and W3 actually
represents the deviation of real thermal emissivity relative to the model-input emissivity
$\approx0.9$, while the ratios of observation/model for band W1 stand for deviation of real
geometric albedo from the best-fit value $\approx0.064$.
\begin{figure*}[htbp]
\includegraphics[scale=0.58]{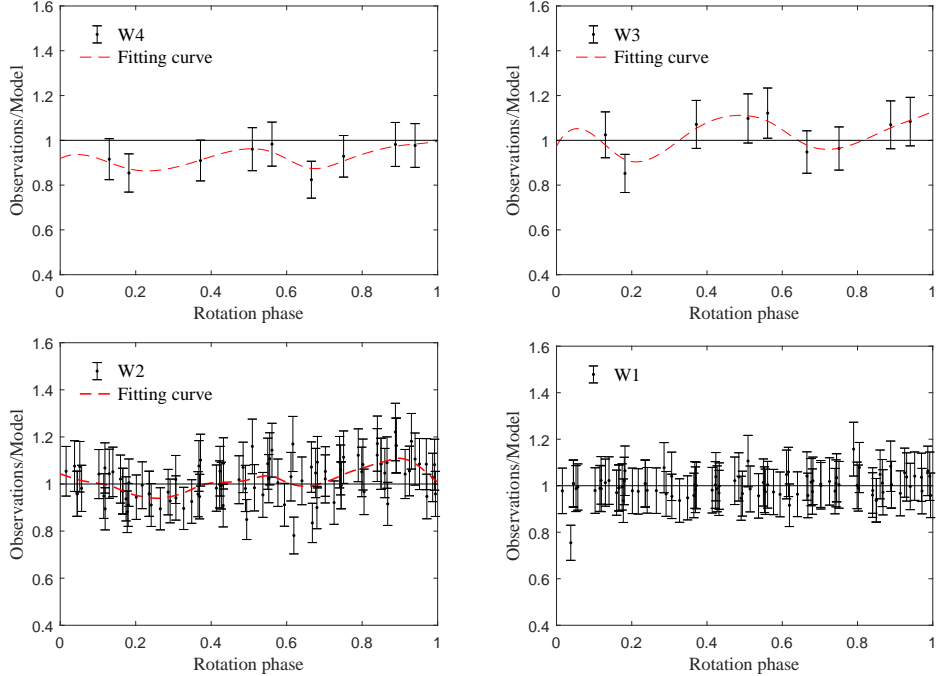}
  \centering
  \caption{Ratios of observation/model as a function of rotation phase for each band respectively.
  }\label{nthlc}
\end{figure*}

Therefore, the ratios of observation/model for each band are separately plotted as
a function of rotation phase in Figure \ref{nthlc}. From Figure \ref{nthlc}, we
see that the observation/model ratios for band W1 are evenly distributed around 1.0
at all rotation phases, showing no rotation-phase dependent feature, hence indicating
that the light-curve inversion shape model fits the W1-band data pretty good and surface
albedo of Themis doesn't significantly vary with longitude. However, for band W4, W3
and W2, the observation/model ratios show a weak rotation-phase dependent feature,
as shown by the red dashed-curves in Figure \ref{nthlc}, where variation trend
of the three red dashed-curves are similar to each other but different from the
trend of W1, indicating that surface materials on different longitude of Themis may
have heterogeneous thermal properties if the shape model imperfection is small.
However, the light-curve inversion shape model is only an ideal shape that achieves
optimum fitting to the visible lightcurves. It may be still different to the real shape,
and hence may not be able to produce thermal light-curve very well. Therefore, the
possibility of trends in W2, W3, and W4 thermal lightcurves being caused by shape
model imperfections cannot be removed.

\section{Discussion and Conclusion}
Thermophysical modelling is the basis of thermal infrared radiometry, which is the
main method to measure thermophysical properties of surface materials on small bodies.
So developing advanced thermophysical models have been always the direction of work
in this field.
For airless small bodies, the thermal state of the surface layers is influenced not
only by the surface thermophysical parameters, but also by geometric effect like
roughness or topography, and kinestate including rotation and orbital motion.
The complexity of the problem lies in that these effects' influence on the thermal
state are coupled together, making that surface thermal parameters and roughness
have inevitable degeneracy in the radiometry procedure, hence multi-epoch
observations are necessary to remove the degeneracy of thermal inertia and roughness
in the radiometric procedure. Therefore, for the use of interpreting multi-epoch
thermal lightcurves (e.g WISE/NEOWISE), we propose this thermophysical model for
realistic surface layers on airless small bodies --- RSTPM, which
simultaneously considers real orbital cycle, rotation cycle, rough surface,
temperature dependent thermal parameters, as well as contributions of sunlight
reflection to observation.

When we aim to interpret multi-epoch observations, the temperature dependence
of thermal inertia becomes non-negligible if seasonal effect can cause
significant temperature variation. As shown in Figure \ref{GammaSv}, seasonal
temperature variation's influence on thermal inertia can be more significant
than the influence from the uncertainties of mean dust-grain size, if there is
a dust mantle on the surface, and thermal conductivity of the dust mantle can
be related to the temperature, porosity of the dust mantle and mean dust-grain
radius like Equation (\ref{ktrphi}). Therefore, for such small bodies covered
by dust mantle, the mean dust-grain size of dust mantle is more suitable to be
used as the free parameter to be determined from the radiometric procedure,
which then provides a way to study the physical properties of dust on the
surface of small bodies.

Of course, there are also small bodies that don't have dust mantle on the
surface. For example, in-situ observations of (162173) Ryugu by Hayabusa2 show
that most of Ryugu's surface is covered by porous boulders \citep{Okada2020}.
For such small bodies, there is even no dust mantle on the surface, hence there
is no need to study dust properties, and thermal inertia can not be modelled as
a function of mean dust grain size, whereas RSTPM can be still used
to study the mean thermal inertia of the surface, or the macroscopic porosity
of the surface if the thermal inertia can be described as a function of
temperature and porosity.

Dust mantle is more likely to appear on large main belt objects, so WISE/NEOWISE
becomes a versatile archive to study dust properties of these bodies.
However, as shown in Figure \ref{rlfraction}, the W1-band observation is actually
dominated by sunlight-reflection, and even the W2-band observation contains
non-negligible $\sim10-30\%$ sunlight-refection. Thus, a precise combination model
of thermal emission and sunlight reflection is extremely necessary to interpret
multi-epoch thermal light-curves of WISE/NEOWISE. RSTPM perform very well
in simultaneously simulating the thermal emission and sunlight reflection, as
demonstrated by its successful application to (24) Themis. But it should be noted
that the "scattering weight-factor" $w_{\rm f}$ used in the reflection model
(Equation \ref{ScatCoeff}) is an artificial factor, which doesn't have clear physical
significance, and may need further examination by more observations and researches.

Nevertheless, the successful application of RSTPM makes it possible for us
to claim that this model is reliable, and is highly capable to derive the physical
properties of small bodies by interpreting the four-band WISE/NEOWISE observations
obtained at multiple epochs. We thus propose: if used to fit rotationally averaged
observations of multiple epochs, RSTPM can study the spin orientation
as well as surface properties, including geometric albedo, roughness and mean
thermal inertia or mean dust-grain size; if used to fit thermal light curves,
RSTPM can investigate whether surface materials on different longitude
are heterogeneous in terms of thermophysical properties.

\section*{Acknowledgments}
We thank the NASA-WISE teams for providing public data. This work was supported by the grants from The Science and Technology Development Fund, Macau SAR (No. 119/2017/A2, 061/2017/A2 and 0007/2019/A) and Faculty Research Grants of The Macau University of Science and Technology (program no. FRG-19-004-SSI).

\bibliographystyle{named}

\end{document}